\documentclass[]{emulateapj}
\usepackage{apjfonts}
\usepackage{natbib}
\usepackage{color}
\usepackage{graphicx}
\usepackage{amsmath}

\shorttitle{Post starburst galaxies in local clusters}
\shortauthors{A. Paccagnella et al.}

\begin{document}
\title{OmegaWINGS: The first complete  census of post starburst galaxies in  clusters in the local universe}

\author{A. Paccagnella\altaffilmark{1,2}}
\author{B. Vulcani\altaffilmark{3}}
\author{B. M. Poggianti\altaffilmark{2}}
\author{J. Fritz\altaffilmark{4}}
\author{G. Fasano\altaffilmark{2}}
\author{A. Moretti\altaffilmark{2}}
\author{Y. Jaff\'e \altaffilmark{5}}
\author{A. Biviano\altaffilmark{6}}
\author{M. Gullieuszik\altaffilmark{2}}
\author{D. Bettoni\altaffilmark{2}}
\author{A. Cava\altaffilmark{7}}
\author{W. Couch\altaffilmark{8}}
\author{M. D'Onofrio\altaffilmark{1,2}}
\affil{\altaffilmark{1}Department of Physics and Astronomy, University of Padova, Vicolo Osservatorio 3, 35122 Padova, Italy}
\affil{\altaffilmark{2}INAF - Astronomical Observatory of Padova, 35122 Padova, Italy}
\affil{\altaffilmark{3}School of Physics, University of Melbourne, VIC 3010, Australia}
\affil{\altaffilmark{4}Instituto de Radioastronom\'\i a y Astrof\'\i sica, IRyA, UNAM, Campus Morelia, A.P. 3-72, C.P. 58089, Mexico}
\affil{\altaffilmark{5}European Southern Observatory, Alonso de Cordova 3107, Vitacura, Casilla 19001, Santiago de Chile, Chile}
\affil{\altaffilmark{6}INAF - Astronomical Observatory of Trieste, 34143 Trieste, Italy}
\affil{\altaffilmark{7}Observatoire de Gen\`eve, Universit\`e de Gen\`eve, 51 Ch. des Maillettes, 1290 Versoix, Switzerland}
\affil{\altaffilmark{8}Australian Astronomical Observatory, PO Box 915, North Ryde, NSW 1670 Australia}

\begin{abstract}
Galaxies that abruptly interrupt their star formation in $<1.5$ Gyr present recognizable  features in their spectra (no emission and H$\delta$ in absorption) and are called post starburst (PSB) galaxies. By studying their stellar population properties and their location within the clusters, we obtain valuable insights on the physical processes responsible for star formation quenching. 
We present the first complete characterization of PSB galaxies in clusters at $0.04<z<0.07$, based on WINGS and OmegaWINGS data, and contrast their properties to those of  passive (PAS) and emission line (EML) galaxies.
For $V<20$, PSBs represent $7.2\pm0.2\%$ of cluster galaxies within 1.2 virial radii. Their incidence slightly increases from the outskirts toward the cluster center and from the least toward the most luminous and massive clusters, defined in terms of X-ray luminosity and velocity dispersion. The phase-space analysis and velocity dispersion profile suggest that PSBs represent a combination of galaxies with different accretion histories. Moreover, PSBs with the strongest H$\delta$ are consistent with being recently accreted. 
PSBs have stellar masses, magnitudes, colors and morphologies intermediate between PAS and EML galaxies, typical of a population in transition from being star forming to passive. 
Comparing the fraction of PSBs to the fraction of galaxies in transition on longer timescales, we estimate that  the short timescale star-formation quenching channel contributes two times more than the long timescale one to the growth of the passive population. Processes like ram-pressure stripping and galaxy-galaxy interactions are more efficient than strangulation in affecting star formation. 
\end{abstract}

\keywords{galaxies: clusters: general - galaxies: evolution - galaxies: formation - galaxies: star formation}

\section{Introduction}\label{intro}

The galaxy population can be thought of as bimodal, with star forming late-type galaxies populating the so-called blue cloud and passive early-type galaxies preferentially found in the red sequence \citep{Blanton2003, Kauffmann2003, Kauffmann2004, Baldry2004,Balogh2004, Brinchmann2004}.
This bimodality is strongly correlated with the environment, with many studies pointing out a well established morphology-density relation holding in both  clusters and the field: the higher the projected surface density of galaxies within an area, the higher is the fraction of early types and the  lower is the fraction of late types \citep[e.g.][]{Dressler1980, Dressler1997, Baldry2004, Fasano2015}. This piece of evidence indicates a role of the environment in shaping galaxy properties. 

The detection of a conspicuous population of star forming galaxies in the core of rich and dense clusters at z$\sim$0.5, but absent today \citep{Butcher1984}, was the first discovery of a strong decline in star formation with time. Subsequently, several studies \citep[e.g.][]{Poggianti1999,Cooper2008} have shown the influence of high density environments
on the formation and evolution of galaxies.  
In an evolutionary scenario in which clusters form and grow through the accretion of field galaxies, these findings suggest that there should be some active mechanisms able to halt the star formation in infalling galaxies.
Several authors proposed many different mechanisms able to shut down star formation in high-density regions; examples include ram-pressure stripping \citep{Gunn1972}, high speed galaxy encounters \citep[galaxy harassment; ][]{Moore1996}, galaxy-galaxy mergers \citep{Mihos1994}, and removal of the warm and hot halo gas \citep[strangulation; ][]{Larson1980, Balogh2000}.
Recently, \citet{Dressler2013} proposed that cluster galaxies could be transformed via group pre-processing, during the fall-in group phase. 

Valuable insights to understand how the star-forming 
population turns passive can be obtained by studying galaxies that appear to have intermediate properties and may be in the act of transitioning between the two main galaxy populations. 
This transition can occur on different time scales depending on the main process inducing the transformations \citep{Wetzel2013, Mok2013, Schawinski2014, Wheeler2014, Vulcani2015, Paccagnella2016}. However, a clear picture describing the reasons why galaxies turn into passive following different paths and the properties of these galaxies is still lacking.  

\citet{Dressler1982}, investigating cluster galaxies at intermediate redhsifts, found a large number of  spectra showing strong Balmer absorption lines and no emission and called them post-starburst galaxies. These features are typical of stellar populations in which star formation had ended abruptly within the last 1-1.5 Gyr and therefore should have been affected by some of the aforementioned processes.

Since then, extensive spectrophotometric modeling \citep{Couch1987,Newberry1990,Poggianti1996,Abraham1996, Poggianti1997,Bekki2001,Poggianti2004} has found that the presence of strong hydrogen lines in absorption and the concomitant absence of emission lines, indicating no ongoing star formation, can be explained roughly decomposing the spectra into a combination of K giant star (or early type galaxy) spectrum and an A star spectrum, Such decomposition is at the origin of the name "k+a", often used to describe post-starburst galaxies. 
In general, A type stars - i.e. new born stars formed within the last 1 Gyr whose spectra are characterized by strong Balmer absorption lines - dominate the light of a galaxy about 0.5 Gyr after star formation stopped and leave their signature on the spectra visible for 1-1.5 Gyr. In contrast, O and B stars - more massive stars that die very quickly and emit the energetic photons able to ionize the gas and produce emission lines - have no or very little contribution.
This combination can appear as a result of a star-bursting episode observed shortly after the star formation has stopped (for those galaxies with the strongest Balmer lines), or as a consequence of normal star formation that has been abruptly ended. Moreover, since A type stars have known lifetimes, the evolution of this population can be used as a quenching clock.

The reason why post-starburst galaxies underwent an episode of star formation that was abruptly stopped is still matter of debate; any of the mentioned quenching mechanisms acting on a short time scale could trigger a starburst and cease star formation, producing a k+a spectra. 
Important hints for solving this matter come from the study of the galaxy environments as different quenching mechanisms play different roles in different environments.

Galaxy mergers could be the dominant mechanism in the  field at low-z, where recent studies found a significant fraction of post starburst galaxies \citep{Bekki2001, Quintero2004, Blake2004,Goto2005, Hogg2006,Mahajan2013}, but they are less effective in clusters, due to the high velocity dispersions.

The origin of k+a galaxies in clusters is more probably related to the interaction of the infalling population with the hot and dense intra cluster medium (ICM).
Indeed, as suggested by \citet{Dressler1982, Couch1987, Dressler1992, Poggianti1999,Tran2003,Tran2004,Tran2007, Poggianti2009}, the interaction of a gas rich galaxy with the hot high-pressure ICM via ram-pressure stripping might trigger a starburst and then clear the disk of neutral gas stopping star formation. 
The most striking evidence for this is perhaps in the Coma cluster, where the position of young post-starburst galaxies is strongly correlated with strong X-ray temperature gradients \citep{Poggianti2004}.
While this picture is corroborated by observations of high and intermediate redshift clusters \citep{Poggianti1999,Muzzin2014}, in the local Universe several works \citep[e.g.][]{Blake2004, Goto2005} found that post starburst galaxies prefer low density environments and thus asses that cluster specific processes are not likely to be the dominant source of fast quenching. Nonetheless, the few studies that investigated the fraction and properties of the local post-starburst cluster population \citep[e.g.][]{Caldwell1997,Poggianti2004, Fritz2014} found evidence of truncated star formation in a significant fraction of cluster members. 
A complete census of post-starburst galaxies in clusters and a homogeneous comparison with the field population is however still lacking. The main reasons are the paucity of the cluster samples studied and the different selection criteria adopted in these studies that do not allow fair intra-sample comparisons. In addition, local post starburst galaxies are often selected without constraints on the H$\alpha$ line, to allow for direct comparisons with high redshift samples, where generally the spectral coverage does not allow to reach the H$\alpha$ emission. However, as shown by \citet{Goto2003} and \citet{Blake2004}, such selections suffer from high levels ($\approx 50\%$) of contamination from H${\alpha}$ emitting galaxies.

However, most E+A galaxies at low redshift are located in the field and represent a very low fraction of the overall galaxy population (0.2 per cent; Zabludoff et al. 1996). The rarity of E+A galaxies at low redshift means that samples can only be constructed from large galaxy redshift surveys. As a result, very few local (v < 5000 km s−1) examples of field E+As are known.

In this work, we exploit a sample of galaxies in clusters drawn from the WIde-field Nearby Galaxy-cluster Survey (WINGS) \citep{Fasano2006, Moretti2014}, and OmegaWINGS surveys \citep[]{Gullieusik2015, Moretti2017}. The combination of the two projects  allows us to study the properties of $\sim 10000$ member galaxies in an homogeneous sample  at $0.04<z<0.07$.
This dataset is unique, as none of the other low-z surveys investigate a large sample of clusters and cluster galaxies in such detail. Thanks to the wide area covered by OmegaWINGS ($\sim 1 deg ^2$), we can investigate cluster members well beyond the cluster virial radius and link clusters with the surrounding population and the field that have been proved to be essential for understanding galaxy transformations  \citep[][]{Lewis2002, Pimbblet2002, Treu2003, Moran2007,Marziani2016}. 

We investigate the occurrence and properties of post starburst galaxies in 32 clusters and compare them to those of passive and star forming galaxies. 
Trends are investigated  as a function of both clustercentric distance and  global cluster properties, i.e. cluster velocity dispersion and X-ray luminosity. Aim of this analysis is to shed light on the processes acting on these galaxies and the time scale needed to transform from one type to the other.

The paper is organized as follows. Section \ref{sec:dataset} presents the dataset and the main galaxy properties; Sec. \ref{sec:class} summarizes the spectroscopic classification; Sec. \ref{sec:datasample} presents the data sample. Section \ref{sec:results} presents our results, focusing on the analysis of the post starburst galaxy population and comparing it to the complementary samples of star-forming and passive galaxies. Morphologies, spatial distributions, and dependencies on global cluster properties are inspected.
Finally, we discuss our results in Sec. \ref{sec:disc} and conclude in Sec. \ref{sec:conc}. 

Throughout the paper, we adopt a \citet{Salpeter1955} initial mass function in the mass range 0.15-120 M$_{\odot}$. The cosmological constants assumed are $\Omega_m=0.3$, $\Omega_{\Lambda}=0.7$ and H$_0=70$ km s$^{-1}$ Mpc$^{-1}$.

\section{Dataset and galaxy properties}\label{sec:dataset}
We base our analysis on the WIde-field Nearby Galaxy-cluster Survey (WINGS) \citep{Fasano2006, Moretti2014}, a multi-wavelength survey of 76 clusters of galaxies with $0.04<z<0.07$ X-ray selected from ROSAT All Sky Survey data \citep{Ebeling1996,Ebeling1998,Ebeling2000}, and from its recent extension, OmegaWINGS, that includes new observations for 46 of these clusters (\citealt{Gullieusik2015,Moretti2017}). The cluster samples cover a wide range of velocity dispersion ($\sigma_{cl}\sim$500-1300 km/s) and X-ray luminosity ($L_X\sim 0.2-5 \times 10^{44}$ erg/s).

The WINGS survey is mainly based on optical B, V imaging \citep{Varela2009}  that covers a $34^{\prime}\times34^{\prime}$ field of view, corresponding to at least about a clustercentric distance of 0.6$R_{200}$. R$_{200}$ is defined to be the radius delimiting a sphere with interior mean density 200 times the critical density and is used as an approximation for the cluster virial radius. The survey has been complemented by a spectroscopic survey for a subsample of 48 clusters, obtained with the spectrographs WYFFOS@WHT and 2dF@AAT \citep{Cava2009}, by a near-infrared (J, K) survey for a subsample of 28 clusters obtained with WFCAM@UKIRT \citep{Valentinuzzi2009}, and by U broad-band and H$\alpha$ narrow-band imaging for  a subsample  of 17 clusters, obtained with wide-field cameras at different telescopes (INT, LBT, Bok) \citep{Omizzolo2014}.

OmegaWINGS extends the WINGS survey in terms of cluster spatial coverage:  OmegaCAM/VST imaging in the u, B, and V bands have been obtained for 45 fields covering 46 WINGS clusters over an area of $\sim$1~deg$^2$ \citep{Gullieusik2015}, thus allowing us to investigate trends well beyond the virial radius and connect clusters with the surrounding population and the field. The spectroscopic follow-up has been obtained for a subsample of 33 clusters with the  2dFdr@AAT \citep{Moretti2017}.

The target selection was similar for the two surveys \citep[]{Cava2009, Moretti2017}. It was based on the available optical B, V photometry \citep{Varela2009,Gullieusik2015} and aimed at maximizing the chances of observing galaxies at the cluster redshift without biasing the cluster sample.
Targets were selected to have a total magnitude brighter than $V=20$, excluding only those well above the color-magnitude sequence with $B-V > 1.20$. These criteria minimize the contamination from background galaxies and include all galaxies on and below the red sequence up to $z\sim 0.08$.

Combining the data of the two surveys, the final spectroscopic sample consists of 22674 spectra in 60 clusters \citep{Moretti2017}. 

To measure spectroscopic redshifts, we adopted a semi-automatic method, which involves the automatic cross-correlation technique and the emission lines identification, with a very high success rate ($\approx$ 95\% for the whole sample, see \citealt{Cava2009} and \citealt{Moretti2017}. The mean redshift z$_{cl}$ and the rest frame velocity dispersion $\sigma_{cl}$ of each cluster were derived using the biweight robust location and scale estimators \citep{Beers1990} and applying an iterative 3$\sigma$ clipping.
Galaxies were considered cluster members if they lie within 3$\sigma_{cl}$ from the cluster redshift. 
R$_{200}$ was computed from $\sigma_{cl}$ following \citet{Poggianti2006} and used to scale the distances from the Brightest Cluster Galaxy (BCG).

The spectroscopic catalog has been corrected for both geometrical and magnitude incompleteness, using the ratio of number of spectra yielding a redshift to the total number of galaxies in the parent photometric catalog, calculated both as a function of V magnitude and radial projected distance from the BCG. 

Galaxy properties have been derived by fitting the fiber spectra with SINOPSIS (SImulatiNg OPtical Spectra wIth Stellar population models), a spectrophotometric model fully described in \citet{Fritz2007, Fritz2011,Fritz2014}. 
It is based on a stellar population synthesis technique that reproduces the observed optical galaxy spectra. 
All the main spectrophotometric features are reproduced by summing the theoretical spectra of simple stellar populations of 12 different ages (from 3 $\times 10^6$ to approximately 14$\times 10^9$ years).

The code provides estimates of star formation rates (SFRs), stellar masses ($M_*$), both observed and absolute model magnitudes and measures of observed equivalent widths (EWs) for the most prominent spectral lines, both in absorption and in emission. 
Magnitudes were computed by convolving the filters response curves with the spectrum.

Due to the configuration of the 2dFdr spectrograph, which is a dual-beam system with two arms overlapping around 5700 \AA, for each object observed with this instrument we obtained two spectra, hereafter called red and blue, which are spliced into the full final spectrum. Despite the quite extended overlap, the region of the spliced spectrum often results quite noisy, due to normalization issues. 

While none of the lines used in our analysis (see sect.\ref{sec:class}) fall in the spliced region, the SFRs and mass estimates in such region might be significantly affected by noise, therefore, we fit the continuum only on the blue part of the spectra, ranging from about 3600 \AA{} to 5700 \AA.

Hereafter, we will use stellar masses  locked into stars, including both those that are still in the nuclear-burning phase, and remnants such as white dwarfs, neutron stars and stellar black holes.

 We note the SINOPSIS is fed with fiber spectra,\footnote{ Fibers are centered on the source emission profile with an accuracy that is equal or better than 0.1$^{\prime\prime}$ \citep[see][]{Gullieusik2015}.} therefore all the galaxy properties suffer from aperture effects. The fiber diameters were 2.16$^{\prime\prime}$ (AAT) and 1.6$^{\prime\prime}$ (WHT), therefore the spectra cover only the central 1.3 to 2.8 kpc of our galaxies depending on the cluster redshift, (see \citet{Cava2009} for details), which correspond to approximately half the typical effective diameter of WINGS galaxies.\footnote{ Effective radii were computed with GASPHOT by \citet{DOnofrio2014}; the median value of the circularized effective radius is $\sim$1.7$\pm$0.2 kpc.}
To recover the galaxy-wide integrated properties, all the derived quantities have been scaled from the fiber to the total magnitude, using the ratio of total to aperture fluxes.

For the WINGS sample, morphological types were derived from V-band images using MORPHOT, an automatic tool for galaxy morphology, purposely devised in the framework of the WINGS project \citep{Fasano2012}. MORPHOT extends the classical CAS classification using 20 different morphological diagnostics and assigns a morphological type (MORPHOT type, T$_M$) to each galaxy from -6 (cD) to 11 (irregulars). The morphological classification of the OmegaWINGS sample is currently underway (Fasano et al. in prep.). 
For our purposes, one of us (G.F.) visually classified the morphologies of the post starburst galaxies in our sample (see \S\ref{sec:datasample}) that are not in the original WINGS sample, by inspecting the  V-band images. 

In the following, we will consider three main morphological classes: ellipticals ($-5\leq\textrm{T}_M<-4.25$), S0s ($-4.25\leq\textrm{T}_M\leq 0$) and late-types ($\textrm{T}_M>0$).

\begin{table}
\begin{center}
\caption{Cluster sample: global properties}
\label{tab_gp}
\begin{tabular}{c c c c c c}
\hline\hline
  \multicolumn{1}{c}{Cluster} &
  \multicolumn{1}{c}{z} &
  \multicolumn{1}{c}{N$_{mem}$} &
  \multicolumn{1}{c}{$\sigma_{cl}$} &
  \multicolumn{1}{c}{R$_{200}$} &
  \multicolumn{1}{c}{log(L$_X$)}\\
  \multicolumn{1}{c}{ } &
  \multicolumn{1}{c}{ } &
  \multicolumn{1}{c}{ } &
  \multicolumn{1}{c}{km s$^{-1}$} &
  \multicolumn{1}{c}{Mpc} &
  \multicolumn{1}{c}{10$^{44}$ erg s$^{-1}$} \\
\hline\\
  A1069 &	 0.0651 &	130 &	695  $\pm 55$ &    1.67 &  43.98\\
  A151 &	 0.0538 &	248 &	738  $\pm 32$ &    1.78 &  44.0 \\
  A1631a &	 0.0465 &	369 &	760  $\pm 29$ &    1.84 &  43.86\\
  A168 &	 0.0453 &	141 &	547  $\pm 38$ &    1.32 &  44.04\\
  A193 &	 0.0484 &	101 &	764  $\pm 58$ &    1.85 &  44.19\\
  A2382 &	 0.0639 &	322 &	698  $\pm 30$ &    1.67 &  43.96\\
  A2399 &	 0.0577 &	291 &	730  $\pm 35$ &    1.75 &  44.0 \\
  A2415 &	 0.0578 &	194 &	690  $\pm 38$ &    1.66 &  44.23\\
  A2457 &	 0.0587 &	249 &	680  $\pm 37$ &    1.63 &  44.16\\
  A2717 &	 0.0498 &	135 &	544  $\pm 47$ &    1.32 &  44.0 \\
  A2734 &	 0.0618 &	220 &	781  $\pm 49$ &    1.88 &  44.41\\
  A3128 &	 0.0603 &	480 &	839  $\pm 29$ &    2.02 &  44.33\\
  A3158 &	 0.0594 &	357 &	1024 $\pm 37$ &    2.46 &  44.73\\
  A3266 &	 0.0596 &	678 &	1319 $\pm 40$ &    3.17 &  44.79\\
  A3376 &	 0.0463 &	263 &	845  $\pm 42$ &    2.04 &  44.39\\
  A3395 &	 0.0507 &	369 &	1206 $\pm 55$ &    2.91 &  44.45\\
  A3528 &	 0.0545 &	262 &	1017 $\pm 46$ &    2.45 &  44.12\\
  A3530 &	 0.0549 &	275 &	674  $\pm 39$ &    1.62 &  43.94\\
  A3556 &	 0.048  &	359 &	669  $\pm 35$ &    1.62 &  43.97\\
  A3558 &	 0.0486&	442 &	1003 $\pm 34$ &    2.42 &  44.8 \\
  A3560 &	 0.0491 &	283 &	840  $\pm 35$ &    2.03 &  44.12\\
  A3667 &	 0.0558 &	386 &	1011 $\pm 42$ &    2.43 &  44.94\\
  A3716 &	 0.0457 &	327 &	849  $\pm 27$ &    2.05 &  44.0 \\
  A3809 &	 0.0626 &	244 &	554  $\pm 38$ &    1.33 &  44.35\\
  A3880 &	 0.058  &	216 &	688  $\pm 56$ &    1.66 &  44.27\\
  A4059 &	 0.049  &	229 &	752  $\pm 38$ &    1.82 &  44.49\\
  A500 &	 0.0682 &	227 &	791  $\pm 44$ &    1.89 &  44.15\\
  A754 &	 0.0545 &	338 &	919  $\pm 37$ &    2.22 &  44.9 \\
  A85 & 	 0.0559 &	172 &	982  $\pm 55$ &    2.37 &  44.92\\
  A957x &	 0.0451 &	92 &	640  $\pm 47$ &    1.55 &  43.89\\
  A970 &	 0.0589 &	214 &	844  $\pm 49$ &    2.03 &  44.18\\
  IIZW108 &	 0.0486 &	171 &	612  $\pm 38$ &    1.48 &  44.34\\
\hline\end{tabular}
\\
\end{center}
\tablecomments{Columns: (1) cluster name, (2) cluster mean redshift, (3) number of member galaxies (used to compute mean redshift and velocity dispersion as explained in the text), (4) cluster velocity dispersion with errors, (5) R$_{200}$ in Mpc, (6) logarithm of the X-ray luminosity (from \citet{Ebeling1996}). }
\end{table}

\section{The spectral classification}\label{sec:class}

Focus of this paper is the characterization of the properties of galaxies showing different features on their spectra, therefore we rely on the measure of the EWs provided by SINOPSIS. We convert the observed EWs in rest frame values, simply dividing the measurements by  $(1+z)$ and we adopted the usual convention of identifying emission lines with negative values of the EWs and absorption lines with positive ones.

\citet{Fritz2007} found that both absorption and emission lines are reliably measured by SINOPSIS in spectra with S/N$>$3, calculated across the whole spectral range. In spectra with S/N$<$3, noise can be misinterpreted as an emission. This happens especially for the [OII] emission line, which is located in a spectral range where the hydrogen lines of the Balmer series (in absorption) crowd. It is hence possible that a peak between two such absorption lines is misinterpreted as an [OII] emission. In spectra with $S/N\leq 3$, the EW of the [OII] is $\geq -2$\AA, and we use this value as threshold above which measurements are considered unreliable.
In our sample, the average S/N, calculated for the whole spectral range, is $\sim12$.  $2\%$ of the spectra have $S/N<$3, therefore they are only a negligible fraction.

Based on the spectral classification originally proposed by \citet{Dressler1999} and \citet{Poggianti1999} and more recently updated by \citet{Fritz2014}, we subdivide our sample into three classes, according to the rest frame EWs of [OII] and H$\delta$, which are good indicators of current and recent star formation. When the [OII] is not detected, the equivalent width of [OIII] and H$\beta$ are also used.
Differently from \citet{Fritz2014}, our spectral classification exploits also the information on the H$\alpha$ line: all spectra showing H$\alpha$ in emission are directly classified as emission line galaxies, regardless of the other lines.  
In this way we obtain a more robust classification, ensuring  that there is no current star formation, in both the passive and post-starburst samples. 
The detailed description and the physical interpretation of this classification is discussed by \citet{Poggianti1999,Poggianti2009} and \citet{Fritz2014}. 
Briefly, spectra with any of the aforementioned emission lines belong to galaxies in which the star formation is taking place, and will thereafter be called emission line galaxies (EML).
Spectra with no emission lines, including H${\alpha}$, are divided based on the strength of H$\delta$: $k$ spectra, normally found in passively evolving elliptical galaxies, resembling those of K-type stars, with weak H$\delta$ in absorption (H$\delta$<3 \AA), and $k+a$ and $a+k$ spectra, displaying a combination of signatures typical of both K and A-type stars with strong H$\delta$ in absorption (respectively $3<$ H$\delta<8 \AA$ and H$\delta>8 \AA$). The former (k-type) will be thereafter called passive galaxies (PAS), the latter (both $k+a$ and $a+k$) post starburst galaxies (PSB).

We note that, among the PSBs, the strength of H${\delta}$ is indicative of the initial condition associated with the main quenching event. Indeed an
 H${\delta}> 6$ \AA{} can be explained only if a burst of star formation involving high mass fractions (10-20\%) happened prior to the sudden quenching. Galaxies showing this feature are caught in an early phase of transition \citep{Goto2004}. 
In contrast, spectra with a moderate H${\delta}$ line could be both the result of the truncation of star formation in a normal star forming galaxy (thus no burst is required) or a late stage of evolution of the proper post-starburst galaxies. Broadly speaking, while all  PSB galaxies with strong H${\delta}$ will later turn into PSBs with moderate H${\delta}$, the opposite is not true. Therefore, we will sometimes discuss also the strong PSB (H${\delta}> 6$ \AA{}, hereafter sPSB) separately. 

The automatic classification has been visually confirmed. Upon inspection, we noticed that in a number of cases the code had mis-identified emission lines (oxygen forbidden lines and H$\beta$), measuring noise rather than real emission. 
The [OII] emission line, by coincidence, is located in a critical region of our spectra, being in the shortest wavelength regime covered by the spectrograph that often result quite noisy \citep[see][]{Smith2004}.
Moreover, we found that, in most cases, the H${\beta}$ and [OIII] emissions were mis-identified in spectra showing only one emission line.

We visually inspected all the spectra where only one emission line was detected and, if necessary, changed the galaxy spectral type. 
Furthermore, we checked all the $k+a$ and $a+k$ candidates and excluded those with undetected emission lines, remeasured the H${\delta}$ EW for those  with an automatic measure  higher than 5 \AA{} or with a comparable uncertainty.

As already discussed in the previous section, we note that due to the limited size of the fiber diameter, we classify galaxies based on the spectra targeting only the central region of galaxies. The aperture bias could in principle lead to a misclassification of galaxies that may have remaining star-formation activity outside the fiber.

\section{DATA SAMPLE}\label{sec:datasample}

One of the main focus of this paper is the occurrence of PSB galaxies as a function of clustercentric distance; in particular we want to investigate the role of the cluster environment  also beyond the virial radius.
We therefore restrict our analysis to the clusters covered also by the OmegaWINGS observations. 
Among these, only the 32 clusters with a global spectroscopic completeness higher than $\approx$ 50\% are used. The final cluster sample is presented in Table \ref{tab_gp}.

\begin{table*}
\caption{Weighted spectral numbers and fractions}
\centering
\begin{tabular}{ l  |c c| c c| c c| c c  }
\hline	
\hline
Galaxy type & \multicolumn{2}{c|}{PAS} & \multicolumn{2}{c|}{PSB} & \multicolumn{2}{c|}{sPSB} &  \multicolumn{2}{c}{EML}\\
& N & \% & N & \% & N & \% & N & \% \\
\hline
Clusters &   8162 (4235) &55.7$\pm$0.4 & 1057 (560) &7.2$\pm$0.2 & 154 (80) & 1.1$\pm$0.3 & 5441 (3029) & 37.0$\pm$0.4\\
Field &   415 (225) & 19.7$\pm$0.8 & 28 (15) &1.3$\pm$0.2 &7 (3) & 0.3$\pm$0.1 & 1667 (923) & 79.0$\pm$0.9\\
\hline
\end{tabular}
\vspace{5pt}
\tablecomments{Weighted Number (raw numbers in brackets) and percentage of the different spectral types for the magnitude-limited sample weighted for spectroscopic incompleteness and considering only galaxies inside 1.2R$_{200}$. The field sample has no radial limits. The proportion of PAS (k), PSB (k+a/a+k), strong PSB (\textbf{the subsample of} PSB with EW(H${\delta}>=$6) and EML galaxies are listed along with binomial errors.}
\label{tab_frac}
\end{table*}

\begin{figure*}
\vspace{-330pt}
\centering
\includegraphics[scale=0.9]{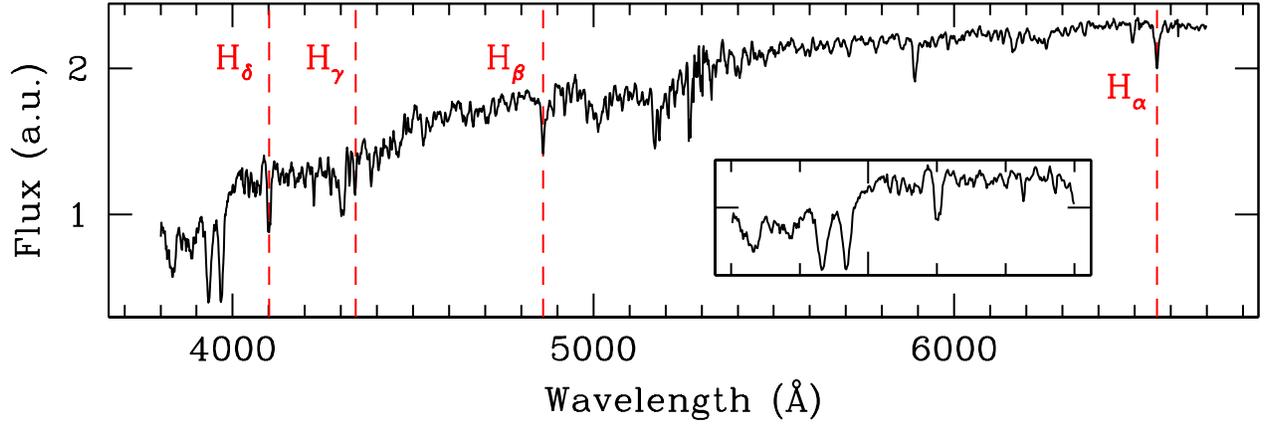}
\vspace{-20pt}
\caption{Rest frame composite spectrum of the post-starburst cluster sample. The inset shows a zoom in of the spectral region around H${\delta}$. \label{comp_spectra}} 
\end{figure*}

Galaxies belonging to the final sample have a V magnitude brighter than 20. 

In our analysis, we neglect the  contribution from Active Galactic Nuclei (AGN). 
As pointed out by \citet{Alatalo2016}, this choice could bias our results in the sense that galaxies in which emission lines are not linked to the star formation process, but are excited by the AGN mechanism, are excluded from the post-starburst sample. 
Nonetheless, \citet{Guglielmo2015}, analyzing a mass limited sample extracted from the original WINGS survey, estimated that the AGN contribution in the star forming galaxy population is approximately 1.6\% and we estimate that this fraction will  not remarkably change in our sample. Thus, including these objects in the emission line galaxy population should not considerably affect our results. A more detailed analysis of emission line galaxies and their classification as star forming galaxies, transition objects and active galactic nuclei for the WINGS sample is presented by \cite{Marziani2016}.

The main sample consists of cluster members within 1.2R$_{200}$ from the BCG, a distance that is reached by almost all the selected clusters (90\%).
We exclude galaxies at larger distances (the maximum radial coverage is approximately 2 R$_{200}$ for member galaxies) because these all belong to clusters with very low velocity dispersion, so they might be a biased population non representative of the general one.  
We also exclude  BCGs, which are a peculiar population \citep[e.g.][and references therein]{ vonderLinden2007,Fasano2010}. 

Non-cluster member galaxies with $0.035<z<0.08$ constitute our field control sample, used for reference.  

The final sample consists of 7824 cluster galaxies and 1163 field galaxies \citep[respectively 14660 and 2110 galaxies, once weighted for incompleteness, see][ for a detailed explanation]{Cava2009}. The number of EML, PAS and PSB galaxies in the different environments is given in Table \ref{tab_frac} and discussed below.

\section{RESULTS}\label{sec:results}
In this section we present our results. We will first investigate the occurrence and the galaxy properties of the different spectral types (e.g. stellar masses, magnitudes, colors and morphologies), then we will investigate trends  as a function of both clustercentric distance and the level of substructures in clusters. 
We will also focus on the position of the galaxies on the phase space and characterize the role of global cluster properties, i.e. cluster velocity dispersion and X-ray luminosity, in driving trends. 
Aim of our analysis is to shed light on the processes that induce a truncation of the star formation on short time scales and give rise to the existence of the PSB galaxies. Given our interest in this population,  in Fig. \ref{comp_spectra} we show the composite spectrum of our PSBs, to visualize the main features of this population. The composite spectrum is obtained by summing the spectra of all the PSB galaxies in the cluster sample, after normalizing each spectrum by its mean value. The measured rest frame EW(H${\delta}$) is greater than 4 {}\AA, the other Balmer lines are well visible and no emission is detected.

\subsection{Properties of the different galaxy populations}
\label{sec:properties}

Table \ref{tab_frac} presents the incidence of the different spectral types in clusters and in the field.  
PAS galaxies dominate the galaxy population in clusters, being 55.7$\pm$0.4\% of all galaxies, while EML galaxies are 37.0$\pm$0.4\% of all members.
In the field, the contribution of the PAS and EML populations is reversed: nearly the 80\% of field galaxies show sign of ongoing star formation, while less than 20\% are PAS.
The fraction of PSB galaxies  is significantly higher  in clusters than in the field: 7.2$\pm$0.2\% vs 1.3$\pm$0.2\%.
We note that our field sample might actually be biased towards galaxies belonging to filaments or structures falling into the main cluster, but we are not able to separate them. 
Furthermore, there are only 15 PSBs in the field, therefore the statistics in this environment is too poor to draw any conclusion. 
A more complete analysis of PSB galaxies in the field in the local universe, based on a different sample, is currently underway (Paccagnella et al in prep.).

\begin{figure*}
\centering
\includegraphics[scale=0.43]{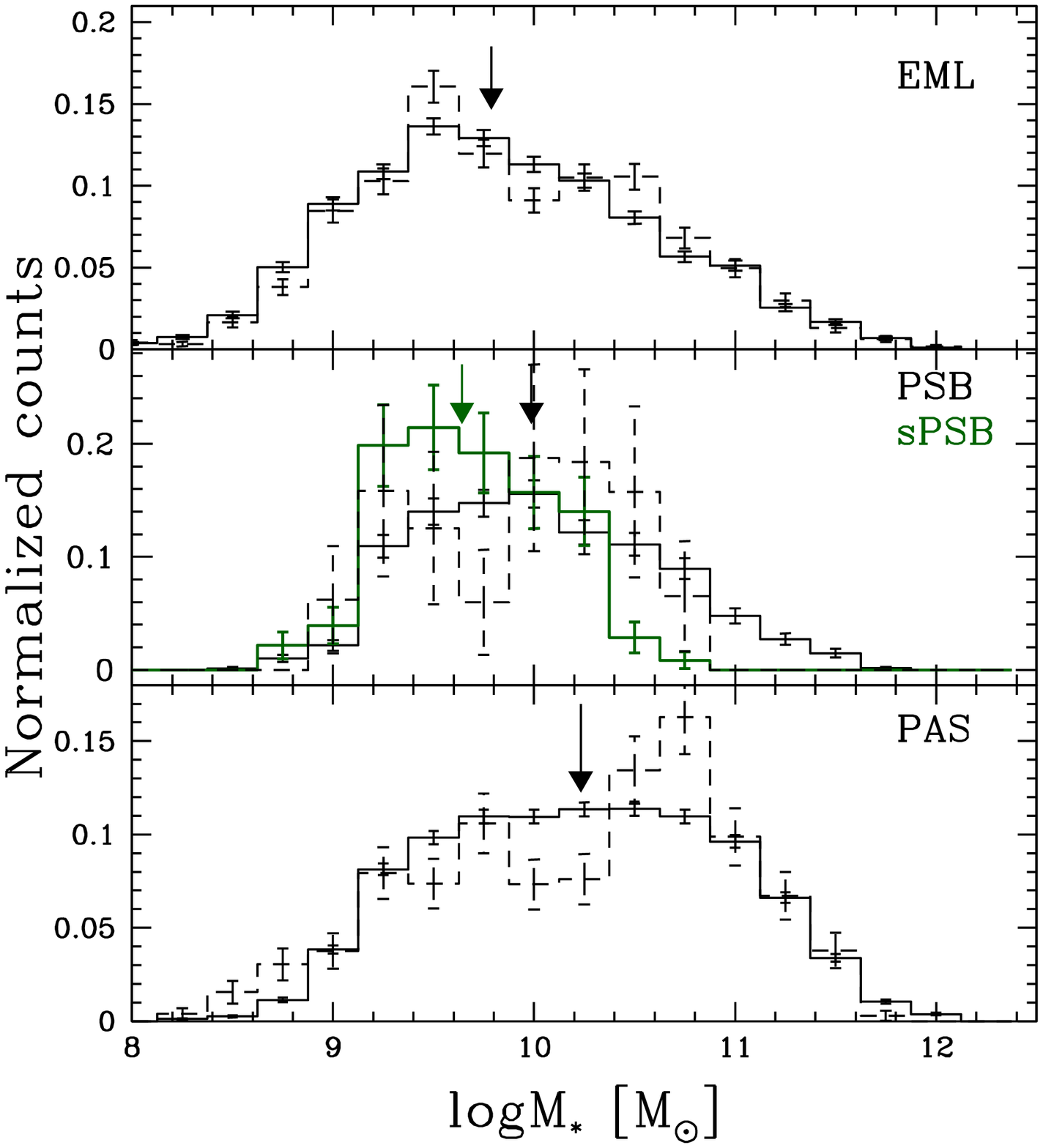}
\includegraphics[scale=0.43]{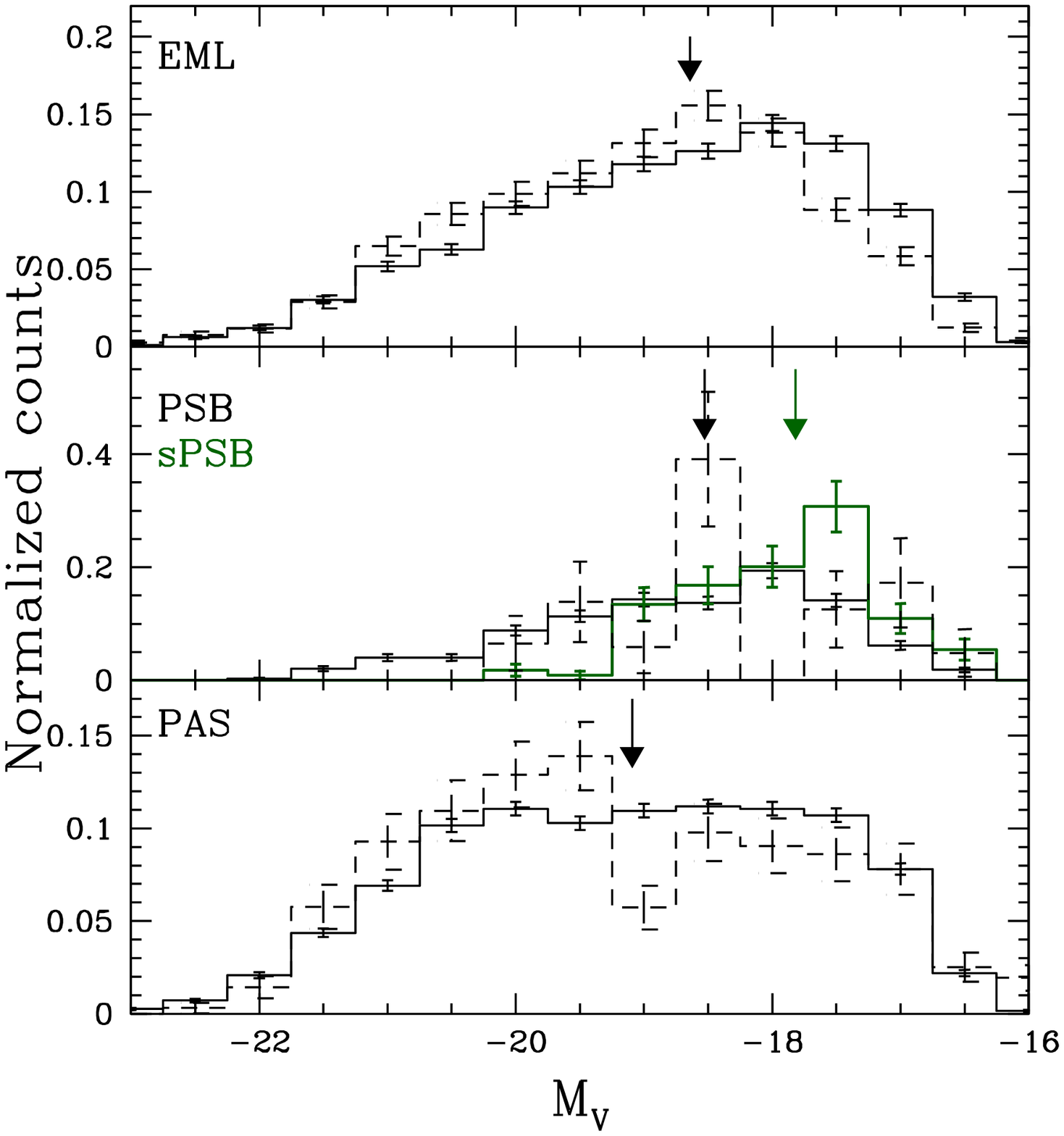}
\caption{Number of cluster (solid) and field (dashed) galaxies in the three main spectral types as a function of stellar mass (left panel) and absolute V magnitude (right panel). Dark green histograms in the middle panels show the sPSB subsample. Arrows indicate the median value for each type. Errors are poissonian \citep{Gehrels1986}. \label{mmdist} }
\end{figure*}

The relevance of the  PSB cluster population is even more striking considering their fraction relative to the active population, which includes galaxies that are (EML) or have been (PSB) star-forming within the last 2 Gyr. The PSB to active fraction  gives the ``quenching efficiency" \citep[see][]{Poggianti2009} that is the efficiency in truncating the star formation in star-forming galaxies. 
PSB galaxies represent 16.3$\pm$0.5\% of the cluster active population, in agreement also with the high redshift fractions derived by \cite{Poggianti2009}, while they make up for less than 2\% in the field.
Hence, clusters are far more efficient than the field in shutting off star formation in galaxies on a very short time scale. 

Stellar mass and absolute V magnitude ($M_V$) weighted distributions for  galaxies of the different spectral types are presented in Fig. \ref{mmdist}. 
The spectral classification, going from PAS to PSB, to EML galaxies, turns out to be, in both environments, a sequence of decreasing mean galaxy mass and increasing mean $M_V$. 
EML galaxies dominate the low mass/low luminosity tails of the distributions, while the contribution of the PAS population becomes  more  important going toward higher masses and luminosities. 
In order to test if the mass and magnitude distributions of the three populations are significantly different, we perform Kolmogorov-Smirnov (K-S) test. The results, comparing the distributions of PSBs to PASs and EMLs  and PASs to EMLs, allow us to reject the null hypothesis that these populations are drawn from the same sample (P-values of the order of 0.0).
This picture fits the  downsizing scenario \citep{Cowie1996}, in which star formation at higher redshifts was more active in more massive/luminous galaxies that are the first to turn into passive.

The distributions of the properties of PSB galaxies are intermediate between those of the PAS and EML populations. This is even clearer looking at the median values of the distributions for cluster galaxies indicated by the black arrows in Fig. \ref{mmdist} (the median masses are 9.78$\pm$0.01, 9.98$\pm$0.02 and 10.23$\pm$0.1, the median magnitudes are $-18.64\pm$0.02, $-18.52\pm$0.04, and $-19.09\pm$0.02 for EMLs, PSBs and PASs, respectively). 
 
Field median values agree, within the standard errors, with the cluster ones, apparently indicating no strong environmental dependencies. However, we remind the reader that the field sample size does not allows us to draw solid conclusions. 

The absolute magnitude distribution of PSBs peaks around M$_{V}=-18.5$, with a deficit among the brightest galaxies (M$_{V}<-20.5$) with respect to both the EML and PAS populations. These characteristics were already visible in the sample described in \cite{Fritz2014} and are similar to those of the post-starburst population in the Coma cluster \citep{Poggianti2004}. 
In the cluster sample, 14$\pm$1\% of the PSB galaxies are classified as sPSB; 
in the field, only 3/15 PSB galaxies are sPSB, making any conclusion statistically meaningless. 

sPSB galaxies (dark green histograms in Fig. \ref{mmdist}) span a narrower range of both absolute magnitude ($M_{V}>-19$) and stellar mass ($\log M_\ast$<10.5) missing the high mass tail and luminosity. This suggests that only the least massive/luminous PSB galaxies undergo a phase of sPSB, and confirms the hypothesis that while all the sPSB galaxies will age and evolve into moderate PSB galaxies, not all PSBs have experienced the sPSB phase.

\begin{figure}
\centering
\includegraphics[scale=0.4]{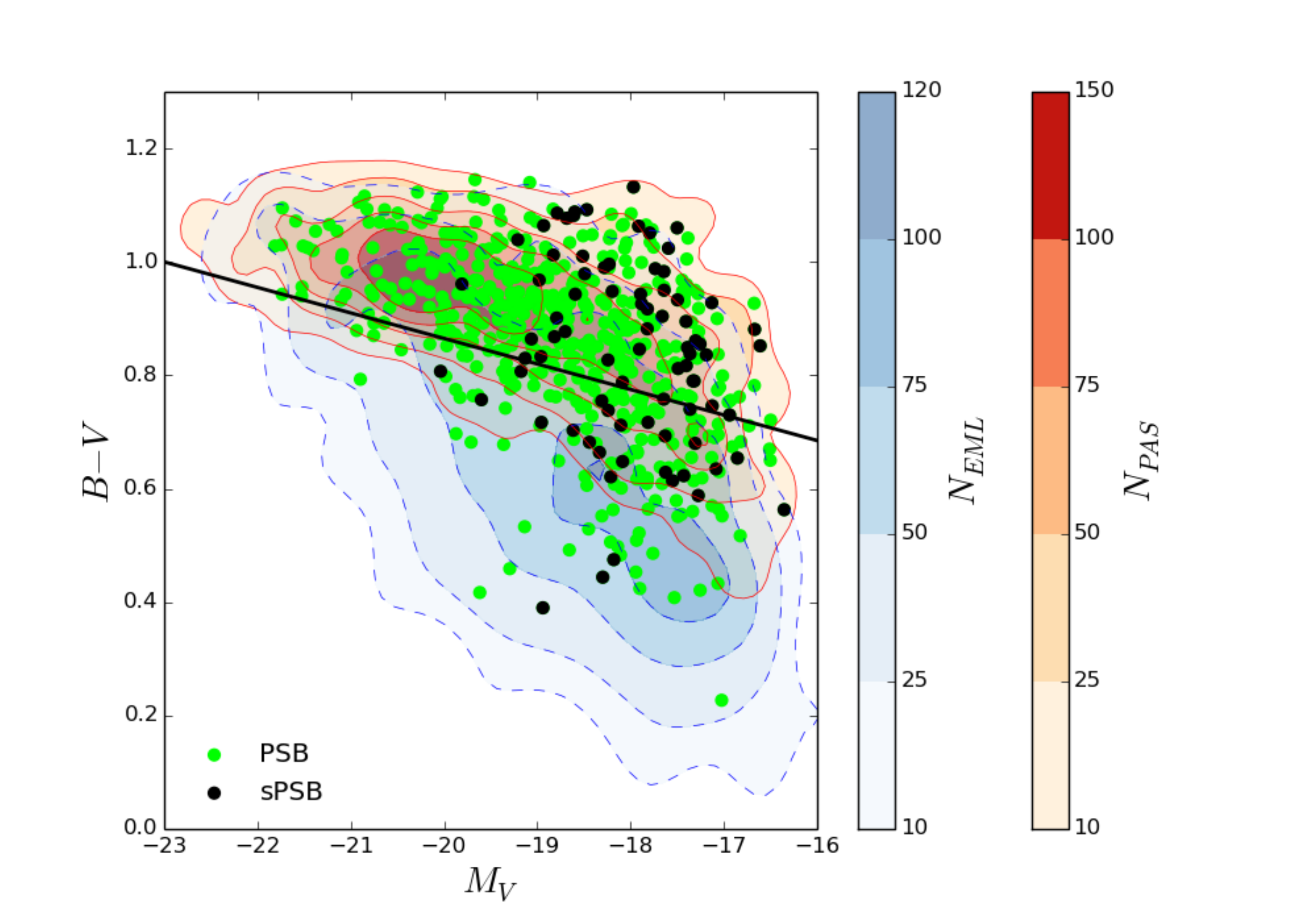}
\caption{Rest-frame ($B-V$)-$V$ relation for cluster galaxies of the different spectral types. Red-solid and blue-dashed contours: number densities of PAS and EML galaxies, respectively. Green dots: PSB; black dots: sPSB. The black line represents the selection limit we adopt to select red and blue galaxies.
\label{fig:col} }
\end{figure}

Figure \ref{fig:col} shows the absolute rest-frame color-magnitude diagram (($B-V$) vs $M_V$) for cluster galaxies of the different spectral classes.\footnote{We recall that absolute magnitudes and rest-frame colors were computed on the fiber spectrum and are given by the spectrophotometric model convolving the filters response curves with the spectrum. } To subdivide galaxies into red and blue, we consider the color-magnitude red sequence of each WINGS cluster  given in  \cite{Valentinuzzi2011}. We use the average value of the slope across all clusters and fix the quote 1$\sigma$ below the average red sequence.  Galaxies whose color lie above 
$$(B-V)_{rf}=-0.045\times V-0.035$$
(black heavy line in Fig. \ref{fig:col}) are assigned to the red sequence, the rest to the blue cloud. 

As expected, most of PAS galaxies present red colors (73.3$\pm$0.5\%), while EML galaxies are preferably blue (75.2$\pm$0.6\%).
Also the majority of PSB and sPSB galaxies have red colors  (73$\pm$1\% and 59$\pm$4\%, respectively), even though the fraction of PSBs with blue colors is not negligible. This finding supports the idea that fast quenching of star formation immediately brings the galaxies to the red sequence, but also that  a pure color-based selection is unable to uniquely distinguish between passive, star forming and galaxies that recently interrupted their star formation. Therefore a detailed spectral analysis is necessary to recognize the different galaxy sub-populations. 

According to the picture we described above, one would expect  sPSB galaxies to have bluer colors than the rest of the PSBs  \citep[see, e.g.,][]{Poggianti2004}. Figure 3 shows that this seems not to be the case in our sample. This results might be due to the fact  that, at low redshift, most of the stars in galaxies are old, and, as soon as the star formation is switched off, the galaxy becomes quickly red as the old stars dominated the integrated light.
An alternative explanation could be the presence of dust.
As already discussed in \cite{Poggianti1999}, the assumption that the progenitors of PSB galaxies are dusty starburst objects entails that dust reddening might affect also the post-starburst class.  
We tested this hypothesis by exploiting the average extinction values given by SINOPSIS and no trend with the EW of  H$\delta$ was found. This result might suggests that either sPSB galaxies are not more obscured than PSBs, or that the real EW values in absence of  dust obscuration should be higher and that our EW estimates have to be considered as lower limits.

\subsection{Morphologies}

\begin{table}
\centering
\caption{Weighted morphological percentages of PSB galaxies}
\begin{tabular}{ l  c c  c }
\hline	
\hline
\vspace{3pt}
 & E & S0 & LT \\
 & \% & \% & \% \\
 \hline
PSB & 28$\pm$1 & 45$\pm$2 & 27$\pm$1\\
sPSB & 17$\pm$3 & 41$\pm$4 & 42$\pm$4 \\
\hline
\end{tabular}
\vspace{5pt}
\tablecomments{Morphological percentages for the PSB and sPSB  samples weighted for spectroscopic incompleteness. Errors are binomial. }\label{tab:morf}
\end{table}

The analysis of the morphologies of PSBs can shed light on the typical properties of this population. 
Table \ref{tab:morf} presents the percentage of galaxies of different morphological types for the whole PSB sample. 
45$\pm$2\% of the galaxies are classified as S0s,  while the remaining sample is evenly divided between ellipticals (28$\pm$1\%) and late-types (27$\pm$1\%).

Considering only sPSB galaxies, while the fraction of S0s does not change within the errors (41$\pm$4\%), that of late types and elliptical does: the former represent 42$\pm$4\% of the total population, the latter 17$\pm$3\%. 

Several reasons might be at the origin of such different morphological distribution between the two samples. 
The larger fraction of late-types among sPSB galaxies than among PSBs agrees with the hypothesis that these are younger objects in which the original disk structure has not yet been changed. 
Together with mass and luminosity distribution of sPSB galaxies (Fig. \ref{mmdist}), this morphological mix could also indicate that the process responsible for the strong burst needed to create the observed spectral features is more effective on less massive galaxies with an high fraction of gas at the moment of infalling.
As an alternative, invoking only ram pressure stripping as the main process responsible for the production of both PSB and sPSB galaxies, the observed morphologies could simply reflect the properties of the infalling population. 
We remind the reader that at low redshifts, the amount of gas necessary to produce a burst of star formation as a consequence of the interaction of the galaxy with the ICM, via ram pressure stripping, is mostly located in low-mass late-type systems while more massive galaxies are mostly early-types and gas deficient \citep{Mahajan2013}.

Similarly to the analysis of galaxy colors, the analysis of the morphologies shows that the majority of PSB galaxies cannot be recognized when using just a morphological classification, but a detailed spectral analysis is necessary.

\subsection{Spatial distribution of the different spectral types}

In the previous sections, we have shown that low-z clusters host a much larger fraction of post-starburst galaxies than the field, pointing towards a cluster-specific origin of the majority of this class of objects.
To differentiate among possible quenching processes that can suddenly truncate the star formation in clusters, we now investigate the radial distribution of the different spectral classes. Overall, the clustercentric distance is a good tracer of the cluster density profile, is related with the time since infall into the cluster \citep{Goto2004}, and
 is  an approximate timescale sensitive to processes occurring on times of the order of a few Gyr. 
Processes which quench star formation gradually, such as stangulation, would induce radial gradients while processes acting on short time scales, e.g. ram pressure stripping, are more likely to cause distinctive signatures at the radii where they are most effective.

\begin{figure}
\centering
\includegraphics[scale=0.43]{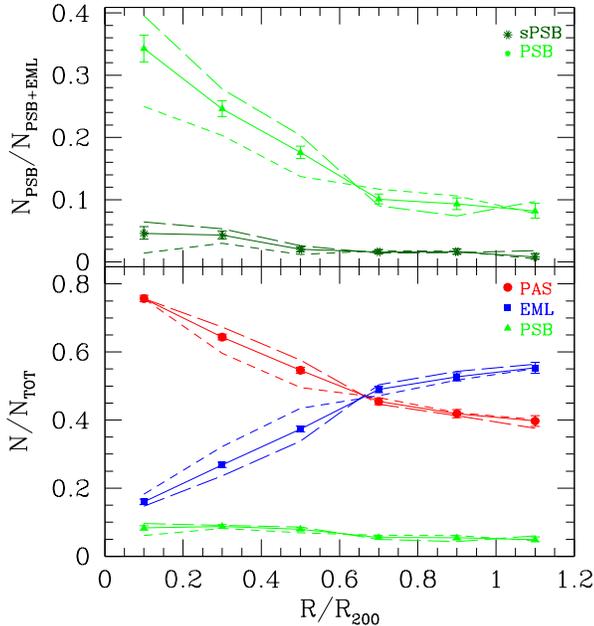}
\caption{Ratio of post-starburst to active galaxies (top panels) and ratio of post-starburst, passive and emission line galaxies to the total (bottom panels) as a function of $R/R_{200}$. 
Points with error bars represent cluster fractions, dashed lines give the trends of the respective fractions in two bins of velocity dispersion $\sigma$ ($\sigma < 840 km/s$ -short dashed lines- and $\sigma > 840 km/s$ -long dashed lines).  Errors are binomial \citep{Gehrels1986}. \label{fig:frac_type}}
\end{figure}

The bottom panel of Fig. \ref{fig:frac_type} shows the incidence of each galaxy population (PAS, EML, PSB) as a function of clustercentric projected distance, in units of R/R$_{200}$. 

As mentioned in Sec. 4, we limit our analysis to 1.2R$_{200}$, not to bias our results towards the clusters with low values of velocity dispersion.
In agreement with previous results, \citep{Weinmann2006,vonderLinden2010, Vulcani2015} we find a pronounced relation between distance from the cluster center and the composition of the galaxy population.
PAS galaxies dominate the inner regions ($\sim$70\% at $R<0.4R_{200}$), and their fraction decreases going outward of a factor of $\sim$2.5. In contrast, the fraction of EML galaxies is $\sim 60\%$ at large clustercentric distances and rapidly declines towards the cluster center of a factor of 4. 
Despite their relatively small incidence, also the fraction of PSB galaxies depends on clustercentric distance, and their trend follows that of PAS: in the cluster cores PSBs are $\sim$1.7 as numerous as PSBs in the outskirts.

The upper panel of Fig. \ref{fig:frac_type} shows the quenching efficiency ($PSB/(PSB+EML)$)  as a  function of the clustercentric distance. 
In the cluster cores, the ratio is $\sim 35\%$, indicating that even in these regions EMLs dominate the active population. The fraction  decreases of a factor of 3 from the cluster center to  0.6R$_{200}$, while it is almost constant in the outer regions ($R/R_{200}$>0.7).
Also the  incidence of sPSBs (dark green symbols in the upper panel of Fig.\ref{fig:frac_type}) among the active population  increases toward the center, even if with a less steep trend.
These results might arise due to the different proportions in the population mix as a function of the global environment, i.e. different cluster halo mass. 
We therefore consider in Fig. \ref{fig:frac_type} two different cluster velocity dispersion bins, respectively higher (long dashed line) and lower (short dashed line) than 840 km/s,\footnote{The value 840 km/s was chosen to approximately divide the galaxy sample in 2 equally populated bins.}
and find that the trends persist and are even more pronounced in the high cluster mass bin. A more detailed analysis of the dependence of the post-starburst fraction on the cluster global properties will be discussed in \S\ref{cl_prop}.

Considering the PSBs of different morphological types, we find that overall the incidence of ellipitcals, S0s and late-types  does not change with clustercentric distance, suggesting there is no morphological segregation for PSBs  (plot not shown).

\subsection{Substructures}
Clusters are generally characterized by the presence of substructures \citep{Ramella2007, Biviano2002}, which implies that merging between clusters and groups is a rather common physical process of cluster formation.
This merging process has been found to affect greatly star formation histories of member galaxies and to eventually induce secondary starburst \citep[see][]{Cohen2014, Bekki1999, Bekki2010}.
The coincidence of the position of the strongest k+a galaxies and the X-ray substructures in Coma found by \cite{Poggianti2004} strengthens this scenario.

\begin{figure*}
\centering
\includegraphics[scale=0.41]{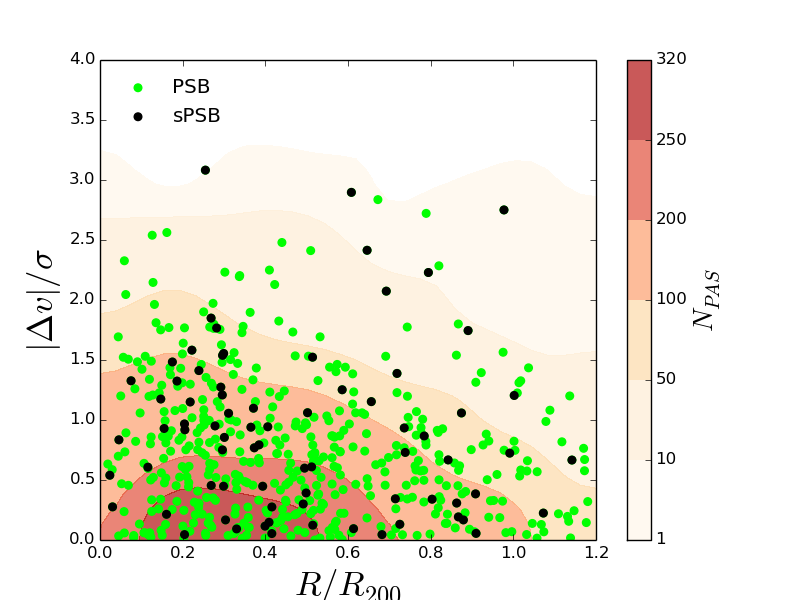}
\includegraphics[scale=0.41]{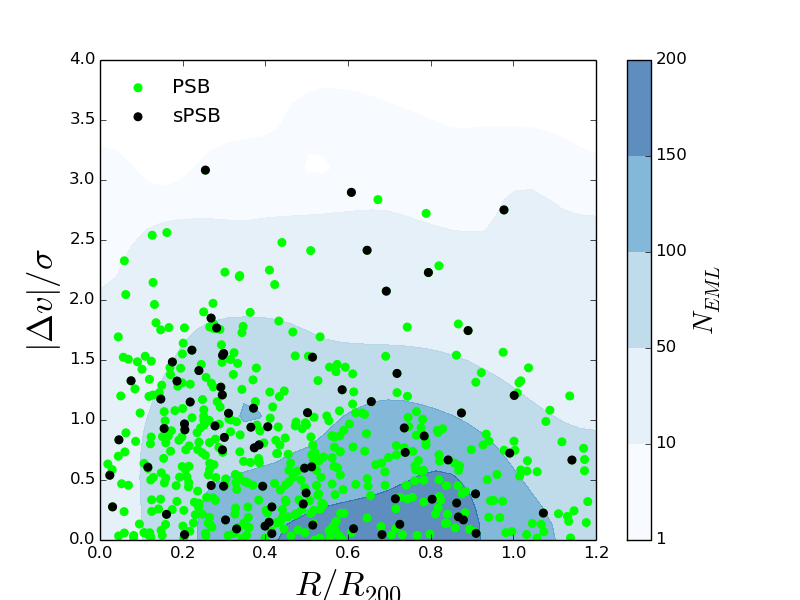}
\caption{Observed stacked phase space diagram for cluster members. Clustercentric distances are in units of R$_{200}$ and the absolute value of the line-of-sight velocities with respect to the cluster recessional velocities are normalized by the velocity dispersion of the cluster. PSB galaxies are represented by green points, sPSB galaxies by black points. They are over-plotted to PAS galaxies (left panel, red contours) and to EML galaxies (right panel, blue contours). \label{pps} }
\end{figure*}

To check whether the presence of sub-structures alters the star formation in cluster galaxies, we compute the fractions of the PSB galaxies belonging to these systems.
We use the catalogs of the WINGS substructures presented by Biviano et al. (in prep.), who define substructures and assign member galaxies to potential substructures exploiting the velocity distributions of galaxies in the cluster local density peaks.

30.4$\pm$0.4\% of the member galaxies belong to substructures and, among these, 6.1$\pm$0.4\% are PSB galaxies and 0.9$\pm$0.2\% are sPSBs. 
The percentage of the PSBs is slightly lower than the value found for the same population when considering the entire galaxy sample (7.2$\pm$0.2\%).
The same conclusion holds for the fraction of PSB and sPSB galaxies with respect to the active population. Thus, this type of analysis does not reveal a strong PSB enhancement in substructures.

We also investigate the fraction of PSB, PAS and EML galaxies in clusters characterized by different levels of relaxation. We use the parameter SUB defined in Biviano et al. (in prep.) that depends, among other things, on the fraction of member galaxies belonging to substructures, and consider three types of clusters: relaxed ($SUB=0$), partly relaxed ($SUB=1$) and unrelaxed ($SUB>1$).
In agreement with e.g. \citet{Cohen2015} and \citet{Biviano1997}, we find a higher/lower fraction of EML/PAS galaxies in less relaxed clusters than in more-relaxed ones (0.344$\pm$0.005/0.580$\pm$0.005; 0.408$\pm$0.0084/0.522$\pm$0.009 and 0.44$\pm$0.01/0.50$\pm$0.01 for SUB=0, SUB=1 and SUB$>1$, respectively).
Even more interestingly, the fraction of PSB galaxies also depends on the dynamical state of the cluster, following the trend of the PAS population: the PSB/PSB+EML fractions are 0.179$\pm$0.006, 0.144$\pm0.009$ and 0.12$\pm$0.01 for SUB=0, 1 and $>1$, respectively.
This is the opposite of what might be expected, if merging clusters were the most favorable environment for PSB production. Part of this trend might be due to the existence of the correlation between PSB fraction and $L_X$ (Fig.~7), as the average X-ray luminosity decreases going from more relaxed to less relaxed clusters (2.8, 2.7 and 1.3 $\times 10^{44} \, \rm erg s^{-1}$ for SUB=0, 1, $>1$). Moreover, the most unrelaxed clusters are in a sense clusters still in the formation process, in which the galaxy populations are still very similar to the unprocessed population of galaxies in the merging groups/clusters, therefore are still very rich of star-forming galaxies that have not experienced a massive cluster environment yet.

\subsection{Phase space analysis}

Many recent papers \citep{Biviano2002, Haines2013,Oman2013,Muzzin2014,Jaffe2016} have shown that galaxy populations with different dynamical histories are well separated in the so-called phase space, which links the spatial position of a galaxy in the clusters, in units of R$_{200}$, and its peculiar velocity, $\Delta v$, normalized by the velocity dispersion of the cluster $\sigma$.
Moreover, the theoretical phase-space diagram derived from cosmological simulations retains information of the epoch of accretion of a galaxy, that can therefore be estimated based on the location of the galaxy in the diagram \citep{Haines2015}.
The same is unfortunately not possible with observations, given the large uncertainties that induce the different populations to overlap on the plane.  
However, it is still possible to retrieve important clues about the dynamical histories of the different galaxy populations in clusters \citep[e.g.][]{Mahajan2011,Oman2013,Muzzin2014,Hernandez-Fernandez2014, Haines2015,Jaffe2015,Jaffe2016}.
Indeed, the overall distribution of galaxies in the phase space strongly depends on the infall times: galaxies that where accreted earlier (i.e. virialized galaxies) typically occupy a triangular-shaped region while the recently accreted or infalling population permeates all projected velocities and radii.
Combining this evidence with the well-established correlation between galaxy quiescence and environment, PAS galaxies are expected to form the majority of the virialized population while EML and PSB galaxies should belong to the infall or recently accreted sample.

Figure \ref{pps} shows the projected phase space obtained combining all clusters together for the different subpopulations separately.
As expected, PAS galaxies are typically found in a triangular-shaped virialized region, which corresponds to the low velocity-low distance area, while EML galaxies are more spread in both radius and velocity.
The PSB population, as already discussed, preferentially lie at small clustercentric distances (R<0.6$R_{200}$), exhibiting anyway a non negligible spread in velocity. 
A 2D K-S test rejects the hypothesis that both PSB and EML and, with a slightly less significance, PSB and PAS galaxies are drawn from the same distribution (P-values respectively of 0.0 and 0.07), suggesting indeed that these populations are in different stages of their virialization process. 
The sPSB population can be better distinguished in terms of velocity rather than clustercentric distance: following the general radial trend of PSB galaxies, these galaxies have slightly higher velocities(median $\Delta v/\sigma$ of 0.67$\pm$0.02 and 0.87$\pm$0.07 for the PSB and sPSB population, respectively). 

\begin{figure}
\centering
\includegraphics[scale=0.4]{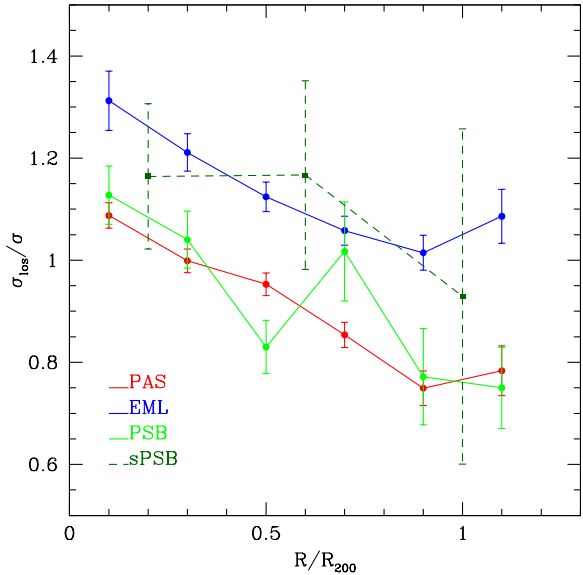}
\caption{Velocity dispersion profiles ($\sigma_{LOS}(r)/\sigma$) of each galaxy population; colors refer to the different populations as described in the labels and as in Fig. \ref{fig:frac_type}. Errors are jackknife standard deviations \citep{Efron1982}. \label{fig:veldisp} }
\end{figure}

To better quantify the differences between the populations, modeling the results presented by \citet{Haines2015} for 75 simulated clusters at z$=0.0$,  Fig. \ref{fig:veldisp} shows the mean normalized line of sight (LOS) velocity dispersion of each sub-population ($\sigma_{LOS}/\sigma$) in six bins of projected radial distance. The errors are obtained using the classical jackknife technique \citep{Efron1982}.
Due to the low number statistics, the sPSB sample is divided only in three bins spanning the same radial range.

By comparison with the analysis of \citet{Haines2015}, the different velocity dispersion profiles can be explained according to the dynamical evolution and accretion history of the galaxy populations. 
PAS and EML galaxies are well separated, the former having low LOS-velocity dispersion at all radii with respect to the latter that displays, especially in the cluster core, higher values of $\sigma$. 
These trends are best reproduced by the virialized population, which was accreted at early epochs ($z>0.4$), and by the most recently accreted and backsplash populations, respectively \citep{Gill2005}
The profile of the PSB population follows the one traced by the PAS$/$virialized population, while sPSBs have remarkably larger velocity dispersions thus belonging to the more recently accreted population.

\begin{figure*}
\centering
\includegraphics[scale=0.4]{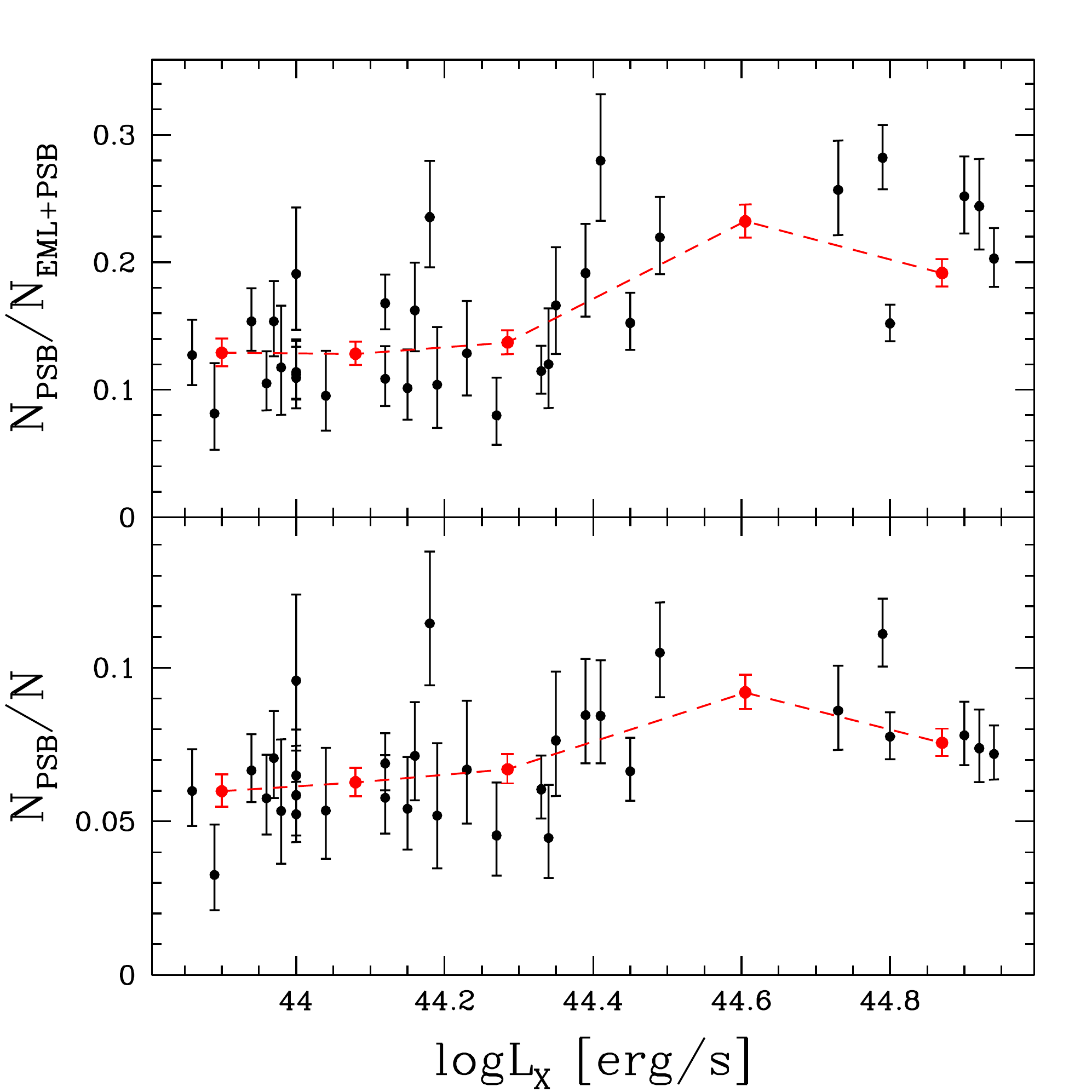}
\includegraphics[scale=0.4]{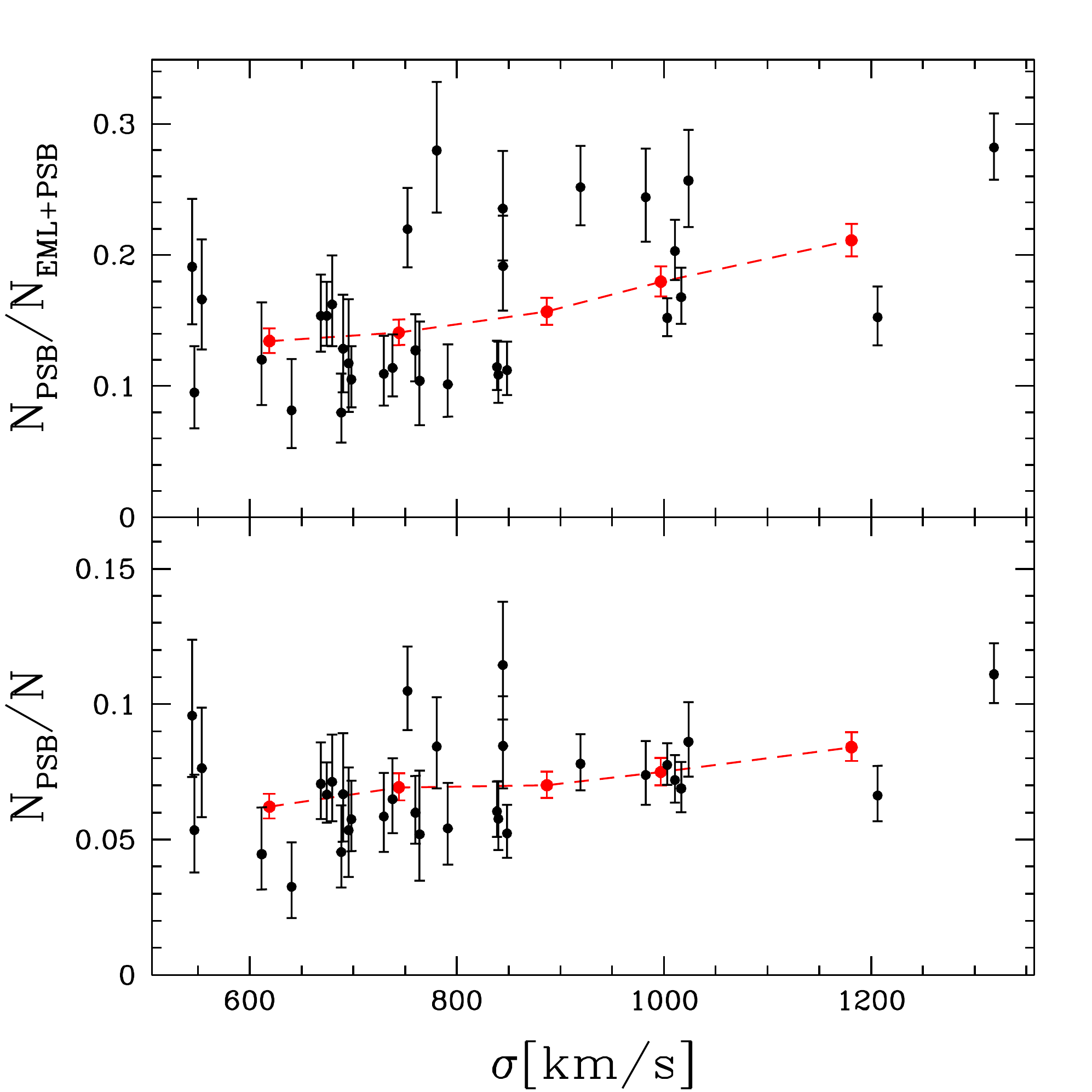}
\caption{Fraction of post-starburst galaxies as a function of the cluster X-ray luminosity (left) and velocity dispersion (right). Black points represent individual clusters, red points give the fractions in 5 equally populated bins. 
Lower panels: weighted fraction of post-starburst galaxies among the whole population. Upper panels:weighted fraction of post-starburst galaxies among the active population. Errors are binomial. \label{fig:fraz_global} }
\end{figure*}

These findings well fit the scenario in which the strength of the EW(H${\delta}$) sets a time line with higher values indicating a more recent quenching event. If the sPSBs are the consequence of a star-burst prior to a sudden quenching, the violent$/$sudden event that triggers the burst seems to happen on first infall; moderate PSBs could originate in the same way but are observed longer time after quenching \citep{Poggianti2009}.

This result moves in the same direction of \cite{Muzzin2014} who, investigating the location of different classes of galaxies on the stacked phase space of clusters at z$\sim$1, found that post-starburst galaxies are commonly found at small clustercentric radii with high clustercentric velocities.

\subsection{Dependence of the spectral type fractions on cluster properties}
\label{cl_prop}

So far we have shown how the fractions of the different subpopulations 
change as a function of the spatial location, both physical and dynamical, within the clusters.  

We now examine if and how the fraction of PSB galaxies depends on the global properties of the clusters, such as velocity dispersion and X-ray luminosity, both proxy for the system total mass.
It is still unclear whether or not the cluster mass affects the amount of observed star formation. For example, several studies at high and intermediate redshifts \citep[e.g.][]{Finn2005,Poggianti2006} find an inverse correlation between star formation and cluster mass, while others \citep[e.g.][]{Popesso2007} assert that no such correlation exists in the nearby cluster population.

All our clusters contain galaxies belonging to all the three main spectral classes, but the population mix in each cluster is quite different.
To assess if these differences arise from a dependence on the cluster halo mass, for each cluster we compute the PSB fraction, weighted for completeness, among all galaxies and among the active population. 
For this analysis, we restricted the data to an absolute magnitude limited sample. This should prevent the introduction of possible systematics related to the selection criteria of the cluster sample.
Member galaxies brighter than $M_V=-17.4$ were selected. This is the absolute magnitude corresponding to the $V=20$ apparent magnitude limit of the most distant cluster.

The results are shown in Fig. \ref{fig:fraz_global} where the individual systems, indicated as black points, are shown as a function of velocity dispersion (left) and X-ray luminosity (right).

To mark the average trends we group the cluster sample into five bins of $\sigma$ and $L_X$ with approximately the same number of galaxies and show the results as red points.
Both fractions (PSB/all and PSB/active galaxies) increase with the mass of the system, more significantly when we consider the X-ray luminosity rather than the velocity dispersion. A Spearman test, which assesses how well the relationship between two variables can be described using a monotonic function, performed on the unbinned data, yields a 99.8\%(99.9\%) and 98.1\%(99.3\%) probability of a correlation of the PSB/all (PSB/active) fraction with L$_{X}$ and $\sigma$ respectively.

This result is in line with the findings of \cite{Poggianti2009} for clusters in the EDisCS sample at $z\sim0.5$ even if the strength of the correlations for our sample is lower, (the Spearman test yields a 99.1\% and 99.7\% probability
of a correlation with the velocity dispersion of their PSB/all and PSB/active fractions, respectively).
In contrast, it is at odds with the analysis performed by \cite{Fritz2014} on the restricted WINGS sample. These authors  did not detect correlations with either the velocity dispersion or the X-ray luminosity. Discrepancies could be due to several reasons, such as the different selection criteria for post-starburst galaxies (they did not use the information on the H$_{\alpha}$ line), the different cut in magnitude, the higher completeness or the larger area covered by our sample.

\section{Discussion}\label{sec:disc}
Exploiting the capabilities of the combined WINGS and OmegaWINGS samples, in this paper we have investigated the properties and the spatial distribution of galaxies that are currently star-forming, that have recently interrupted their star formation and that are already passive in clusters at $0.04<z<0.07$. Our main focus has been on PSB galaxies, to shed light on the processes that induce galaxies to undergo this particular phase during their life.

PSB galaxies are characterized by intermediate physical properties with respect to the EML and PAS galaxies, and are thought to be in a transition phase between these two populations: both their median stellar mass and magnitudes are in between the values found for the other two populations. 
As expected given the fact that they have no ongoing star formation, PSB galaxies present colors similar to the PAS galaxies, even though at faint magnitudes they can also be as blue as the EML galaxies. 

Almost half of the PSBs galaxies have been classified as S0s, while the incidence of elliptical and late-type galaxies depends on the strength of the measured EW of H$\delta$: considering all PSBs, the fraction of ellipitcals and late-types is similar, considering only the sPSB, late-type galaxies dominate the population.  

The fraction of PSB galaxies decreases with increasing distance, suggesting that in the core of the clusters some mechanisms are inducing galaxy transitions. The same fraction also depends on the cluster properties and it steadily increases with increasing $L_x$ and $\sigma$.
Moreover, PSBs do not concentrate as much in the low clustercentric distance-low velocity locus of the phase-space as virialized galaxies do. This, together with the fact that their velocity dispersion is quite intermediate between that of PAS and EML galaxies, especially for the sPSB galaxies, could lead to the interpretation that PSBs are a combination of galaxies with a mix of times-since infall (backsplash + virialized).
Ideally, one should define the virial, infall and backsplash classes following the orbits of the particles using cosmological simulation. This task is beyond the scope of this work but will be addressed in forthcoming papers.

\subsection{Slow and fast quenching mechanisms in clusters}

\begin{figure*}
\centering
\includegraphics[scale=0.42]{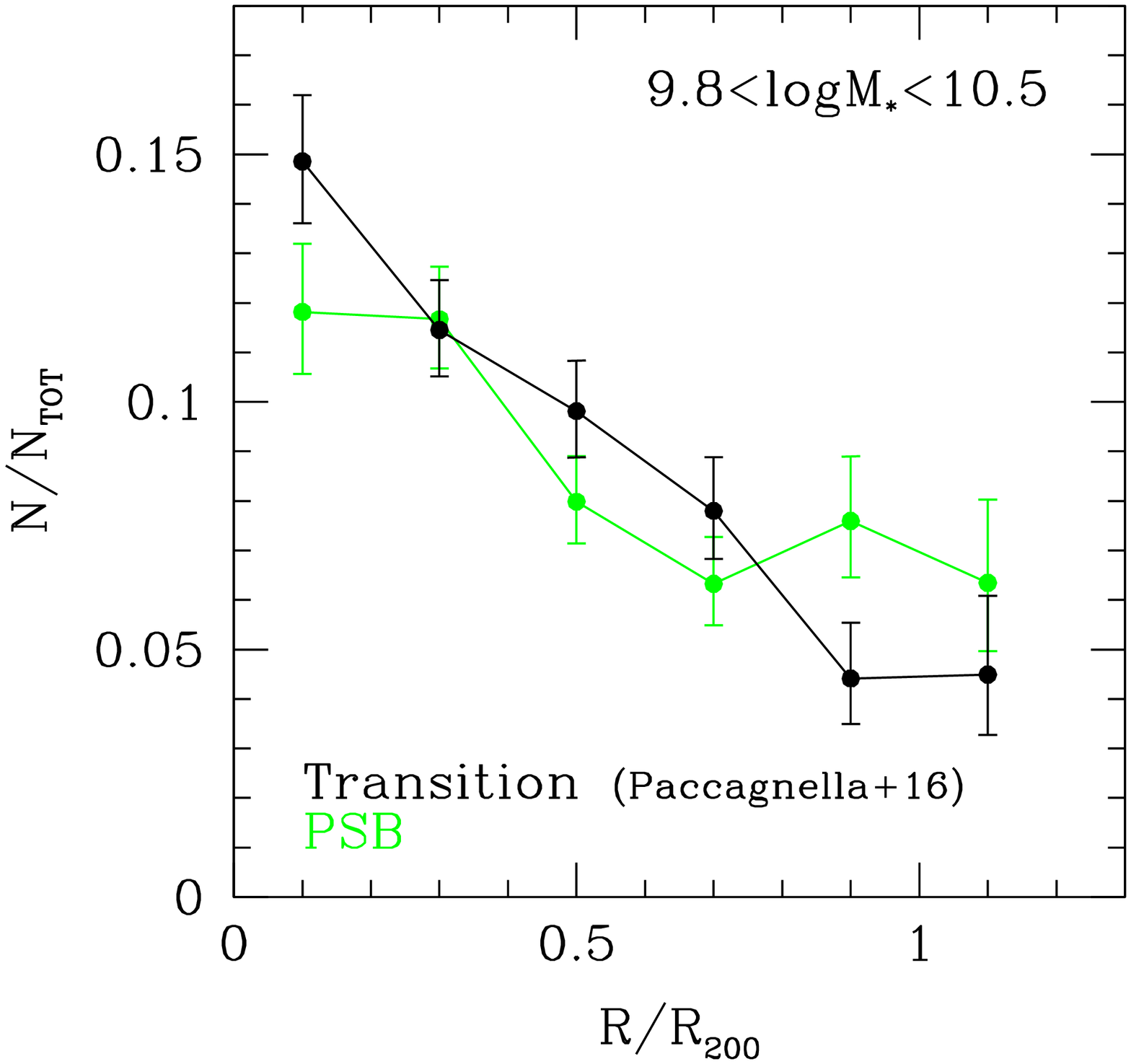}
\includegraphics[scale=0.42]{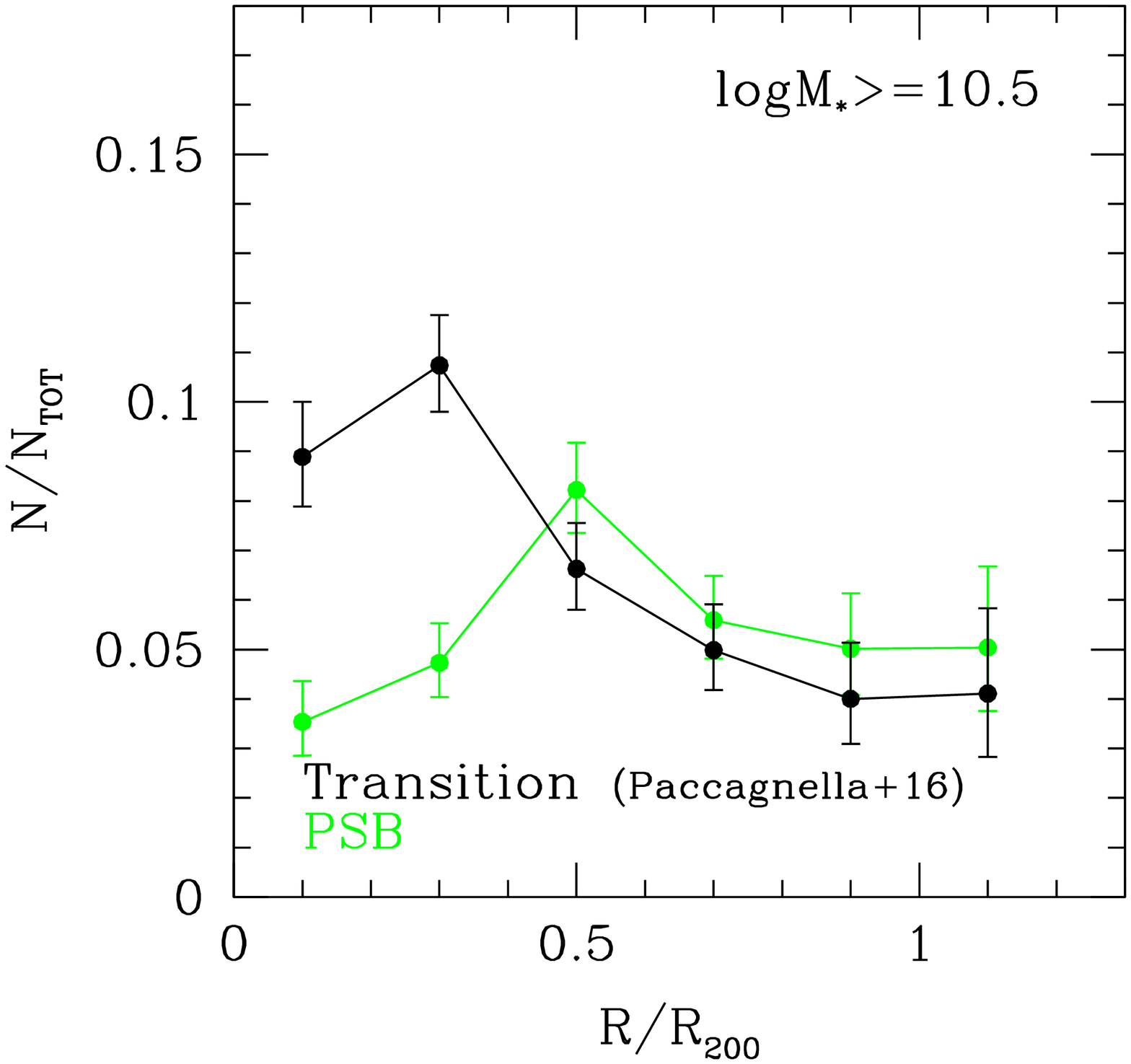}
\caption{Radial distribution of transition galaxies (black) as defined by \cite{Paccagnella2016} and of PSB galaxies (green), in two bins of stellar mass. Errors are binomial.
\label{fig:ka_tr}}
\end{figure*}

As found by \citet{Poggianti2004}, \citet{Tran2007}, and \citet{Muzzin2014}, PSB galaxies have to be generated by a fast acting mechanism, given that this phase can last approximately 1-1.5 Gyr.
As already discusses in Sec.\ref{intro}, spectra showing strong Balmer absorption lines and no emission are the result of a precise combination of an old stellar population above which A-type stars signatures are well visible. These stars, formed within the last Gyr, dominate the light of a galaxy where a recent star formation activity has ended abruptly and are visible for 1-1.5 Gyr \citep[e.g.][and references therein]{Poggianti2004}.
Even though this is a quite efficient channel to transform galaxies from star forming to passive, it is definitely not the only one. E.g. \citet{Patel2009, Vulcani2010, Paccagnella2016} have found a population of galaxies in transition on long time scales (few Gyr) both in the local universe and at higher redshift ($z<1$). Adopting a different approach, following the infall and orbits of galaxies in the vicinity of the 75 most massive clusters in the Millennium cosmological simulation, also \citet{Haines2015} support a slow quenching scenario, with a timescale of the order of 0.7-2 Gyr.

Seeking for objects in transition from being star forming to becoming passive, \citet{Paccagnella2016} analyzed the SFR-M$_*$ relation of local cluster galaxies in a mass-limited sample extracted from the WINGS+OmegaWINGS sample.
They identified a population of galaxies, called \textit{transition galaxies}, located 1.5$\sigma$ below the field SFR-M$_*$ relation, whose star formation histories and properties suggest that they have had a reduced SFR for the past 2-5 Gyr.
At least above the mass completeness limit, the fraction of transition galaxies strongly depends on environment, being almost negligible outside the virial radius and rising towards the center, making up for almost the 30\% of star forming galaxies inside 0.6R$_{200}$.
As concluded by the authors, these findings, together with the estimated quenching time scale, are consistent with the hypothesis that the interaction of galaxies with the intracluster medium via strangulation causes a gradual shut down of star formation. 

We are now in the position of directly comparing the slow quenching channel characterized by \citet{Paccagnella2016} with the much faster one required to observe the typical post-starburst signatures. To do so, we consider only galaxies with stellar mass larger than the mass completeness limit of 10$^{9.8} M_{\odot}$ used by \citet{Paccagnella2016}. 346 PSB galaxies (650 once weighted for completeness) enter the sample,  of which only 32 (56) are sPSB. Above this limit, PSBs constitute the 7.3$\pm$0.3\% of the total population and all the environmental dependencies found for the whole sample persist. 
For comparison, above the same limit and the same clustercentric distance, 408 (780) galaxies have been classified by \citet{Paccagnella2016} as in transition, and they constitute 9.0$\pm$0.3\% of the entire population.
By definition, being transition galaxies selected from the star forming population, there is no object entering both samples.  

Therefore transition galaxies are only slightly more numerous than PSB galaxies. Starting from the logical assumption that both populations have a common progenitor among star-forming galaxies, and if we assume that the transition phase lasts for about twice the time (of the order of $>$2 Gyr) of the PSB visibility ($\sim$1 Gyr), we conclude that the short timescale star-formation ``quenching'' channel contributes at least two times more than the long timescale one to the growth of the passive population. 

Figure \ref{fig:ka_tr} investigates in detail where these two populations are found within the clusters. It shows the number of transition and PSB galaxies to the total number of galaxies, as a function of clustercentric distance. As done by \citet{Paccagnella2016}, two different mass bins are considered here, to also look for trends with stellar mass. 
For $\log M_\ast<10.5$, PSB and transition galaxies present a similar anticorrelation between the fraction of PSB and transition objects and  distance, even though PSB galaxies show a slightly flatter trend, being less numerous than transition galaxies in the core of the clusters and slightly more numerous outside the virial radius. 

For $\log M_\ast>10.5$, at large clustercentric distances, both populations present trends similar to those  at lower masses, but in the cluster cores they present a drop in number. Such drop occurs only in the  cluster cores for transition galaxies, at $r<0.5R_{200}$ for PSBs. Overall, it might be primarily due to the fact that in the cluster cores massive galaxies are already mostly passive, therefore the reservoir for transitioning galaxies is poorer than at lower masses. 
In addition, the different behavior seen for the two populations could be due to the visibility time-scales of the two populations, with PSBs disappearing faster than transition galaxies on the way to the cluster core. 

As mentioned in \S\ref{sec:dataset}, galaxies drawn from the WINGS sample (approximately all those located within 0.6 $R_{200}$), have been morphologically classified. We can therefore compare the morphologies of PSB and transition galaxies in the core of the clusters, to look for signs of a link between galaxy morphology and time scale of quenching. We consider the mass limited samples. 
Overall, among PSB galaxies, 40$\pm$2\% are ellipticals, 44$\pm$2\% are S0s and the remaining 16$\pm$2\% are late-type galaxies. Transition galaxies have a much more numerous population of S0s, with a remarkable 56$\pm$2\% at the expense of elliptical galaxies, accounting only for the 28$\pm$2\%, while 16$\pm$2\% are late-types. 

No strong trends of the  morphological mix with distance have been detected. However, we stress that here we are considering only galaxies in the core of the clusters, and we are not attempting to extrapolate trends at larger distances, where the morphological mix might be different. In addition, we remind the reader that this analysis is performed above the mass completeness limit of 10$^{9.8} M_{\odot}$ and that this cut excludes the vast majority of sPSB galaxies from our sample. This population have been found to have a distinct morphology from the whole population of PSB galaxies (see \S\ref{sec:properties}).

Studies regarding slow and fast quenching should necessarily follow different approaches. 
As far as the post-starburst population is concerned, due to the high precision clock imposed by A-type stars lifetime, there is a general agreement on the time-scales involved \citep[see][ etc.]{ Quintero2004, Poggianti2004, Poggianti2009, Muzzin2014,Vulcani2015} and, from these, different mechanisms have been proposed. 
On the other hand, an hypothesis on the physical processes causing the slow quenching of a galaxy has to be made to estimate the time scales involved.
Several authors have tried to estimate the quenching timescales of satellite galaxies through different approaches. \citet{Wetzel2013} proposed a mass dependent "delay-then-rapid" scenario in which star formation is quenched rapidly but only after a delay of 2-4 Gyr after infall. As an alternative \citet{Taranu2013} presented models where quenching occurs within a smaller radius, approximately $0.5R_{200}$, followed by an exponential decline of star formation over 3-3.5 Gyr. 

\subsection{Specific processes responsible for the objects in transitions in clusters}

We found the majority of transition and PSB galaxies within the virial radius. It seems clear from our results that any of the proposed physical mechanisms (i.e. starvation or ram pressure stripping), which alter the star formation of infalling galaxies, is stronger in the central regions of the cluster, where the density and the temperature of the ICM (as well as the velocity of galaxies) reach their maxima.  
By comparing the cluster crossing times and the A-stars lifetime, we could also imagine an evolutionary sequence in which some of the PSBs (the low mass ones, as discussed in sect. \ref{sec:properties}) descend from sPSB galaxies, i.e. the ram pressure starts being effective at larger distances, generating the population of sPSB, while at smaller cluster-centric distances we only see the PSB galaxies. 
However, the comparison of the radial trend of the fractions of PSBs and transition galaxies does not allow to definitely ascertain whether the two populations have a different origin. Indeed, considering for example that ram pressure stripping depends, among other things, on the orbit and on the orientation of the galaxy with respect to the ICM \citep{Abadi1999}, we can depict two different scenarios in which star formation can either be suddenly shut down (on timescales of the order of $\sim$1Gyr, \citealt{Muzzin2014}) or gently and progressively depressed by ram-pressure stripping, and therefore give origin to the two observed galaxy populations.

In the first scenario, the resulting quenched population is expected to have PSB features, maintaining the original structural properties, since nothing apart from gas loss would disturb its morphology. Recall that the quenching time scale of PSB galaxies is imposed by A-type stars lifetime. 
In the second scenario, ram-pressure stripping might be the main responsible also for the transition population presented in \citet{Paccagnella2016}. Nonetheless, the same population could originate also via strangulation, consistent with a gradual quenching corresponding to an exponential time scale of a 2 Gyr or more. Indeed, strangulation is expected to have the effect of removing the outer galaxy gas halo and prevent further infall of gas into the disk. On timescales of  few Gyr, the star formation would thus exhaust the available gas, quenching the star formation activity.

In addition to ram-pressure stripping and strangulation, other mechanisms might play a role, even though they most likely take place at larger distances form the cluster center.  

As discussed for example by \citet{Treu2003} and reviewed by \citet{Boselli2006}, quenching from gravitational interaction between galaxies (i.e. galaxy-galaxy interactions, harassment) occurs preferentially outside the virial radius, given the high velocity dispersions in the cluster cores that remarkably reduce the probability of pair interactions, with timescales of the order of some $10^{10}$ yr. Moreover, these  interactions act  on the stellar component, producing selective morphological transformations that we do not observe in the transition or PSB populations. 

We emphasize that this does not mean that galaxy-galaxy interactions have no effect at all on cluster galaxies. These processes might eventually become important, but generally after the gas has been removed by other processes.

\section{Summary and Conclusions}\label{sec:conc}
\label{conclusion}
In this work we have resorted to  an observed magnitude limited sample of galaxies in clusters drawn from the WIde-field Nearby Galaxy-cluster Survey (WINGS) \citep{Fasano2006, Moretti2014}, and OmegaWINGS surveys \citep[]{Gullieusik2015, Moretti2017} to investigate the occurrence and the  properties of galaxies of different types in 32 clusters at $0.04<z<0.07$. 
We classified the galaxies according  to the different features detected in their spectra (presence/absence of OII, OIII, H$\delta$ and H$\alpha$) into passive (PAS), post starburst (PSB) and emission line (EML) galaxies. We have compared  stellar population properties and  location within the clusters of the different spectral types to  obtain valuable insights on the physical processes responsible for the star formation quenching. 

The main results can be summarized as follows. 
\begin{itemize}
\item For $V<20$, PAS represent $55.7\pm0.4$\% of the cluster population within 1.2 virial radii, EML represent $37.0\pm0.4$\% and PSBs $7.3\pm0.2\%$, 15\% of which show strong H$\delta$ in absorption (>6, sPSB), indication of either a very recent quenching and/or of a strong burst before quenching.

\item PSBs have stellar masses, magnitudes, colors and morphologies intermediate between PAS and EML galaxies, typical of a population in transition from being star forming to passive. Interestingly,  45\% of PSBs have S0 morphology, $28\pm1$\% are ellipticals and $27\pm1$\% are late types. Considering only sPSBs, the incidence of late-types increases to $42\pm4$\% with a corresponding drops of ellipitcals, which are only $17\pm3$\%. 

\item The incidence of PSBs slightly increases from the cluster outskirts toward the cluster center and from the least toward the most luminous and massive clusters, defined in terms of $L_X$ and velocity dispersion.

\item The dynamical state of the clusters partially influences the incidence of PSBs. While the presence of substructures does not enhance the fraction of PSBs, the level of relaxation does: the fraction of PSBs is higher in relaxed clusters. At least part of this trend is due to the correlation between PSB fraction and $L_X$. 

\item The phase space analysis and the velocity dispersion profiles suggest that PSBs represent a combination of galaxies with different accretion histories. Moreover, the PSBs with the strongest H$\delta$ are consistent with being recently accreted. 

\end{itemize}

PSBs are thought to be galaxies generated by fast acting mechanisms and this phase is expected to last approximately 1-1.5 Gyr \citep[e.g.][]{Poggianti2004, Muzzin2014}. 
Our analysis suggests that as a galaxy is accreted onto a cluster, at first its properties are not strongly affected, but when it approaches the virialized region of the cluster, processes like ram pressure stripping or other interactions induce either a burst of the star formation with a subsequent fast quenching, or simply a fast quenching. As the shut off of the star formation occurs, the galaxy changes the features in its spectrum, but variations in color and morphology require longer time scales, therefore PSBs appear with a wide range of these properties. It is important to stress that the majority of those galaxies that are truncated on a short timescale cannot be recognized based on color or color+morphology, but only by performing a detailed spectral analysis.

The fraction of PSBs is similar to the fraction of galaxies in transition on longer timescales, as defined by \citet{Paccagnella2016}, suggesting that the short timescale star-formation quenching channel, lasting less than half the timescale required to slowly quench star formation, contributes two times more than the long timescale one to the growth of the passive population, therefore processes like ram pressure stripping and interactions are more efficient than strangulation in affecting star formation, at least in clusters. 

In other environments, the fraction of PSB galaxies and the processes responsible for its existence might be considerably different. \cite{Vulcani2015} have found hints of an enhanced fraction of PSB galaxies in groups compared to isolated and binary systems, but a complete characterization of the physical processes is still missing.
In a forthcoming paper (Paccagnella et al. in prep.), we will characterize the incidence of post starburst galaxies in the different environments, contrasting the properties of galaxies in clusters, groups, binary systems and isolated galaxies, to build a complete picture of the assembly of this population.  
 
\section*{Acknowledgments}
We thank the anonymous referee for their useful comments that helped us to improve the manuscript.
AP acknowledges financial support from the Fondazione Ing. Aldo Gini and thanks the Anglo Australian Observatory for a productive stay during which part of this work  was carried out.
We acknowledge financial support from PRIN-INAF 2014 grant.
B.V. acknowledges the support from 
an  Australian Research Council Discovery Early Career Researcher Award (PD0028506).
This work was co-funded under the Marie Curie Actions of the European Commission (FP7-COFUND).


\begin{thebibliography}{}

\bibitem[{Abadi}, {Moore} \& {Bower}(1999)]{Abadi1999}
{Abadi}, M.~G., {Moore}, B., and {Bower}, R.~G. 1999, \mnras, 308, 947.

\bibitem[{Abraham} {\it et al.}\ (1996)]{Abraham1996}
{Abraham}, R.~G. {\it et al.}\  1996, \apj, 471, 694.

\bibitem[{Alatalo} {\it et al.}\ (2016)]{Alatalo2016}
{Alatalo}, K. {\it et al.}\  2016, \apjs, 224, 38.

\bibitem[{Baldry} {\it et al.}\ (2004)]{Baldry2004}
{Baldry}, I.~K. {\it et al.}\  2004, \apj, 600, 681.

\bibitem[{Balogh} {\it et al.}\ (2004)]{Balogh2004}
{Balogh}, M.~L. {\it et al.}\  2004, \apjl, 615, L101.

\bibitem[{Balogh}, {Navarro} \& {Morris}(2000)]{Balogh2000}
{Balogh}, M.~L., {Navarro}, J.~F., and {Morris}, S.~L. 2000, \apj, 540, 113.

\bibitem[{Beers}, {Flynn} \& {Gebhardt}(1990)]{Beers1990}
{Beers}, T.~C., {Flynn}, K., and {Gebhardt}, K. 1990, \aj, 100, 32.

\bibitem[{Bekki}(1999)]{Bekki1999}
{Bekki}, K. 1999, \apjl, 510, L15.

\bibitem[{Bekki}, {Owers} \& {Couch}(2010)]{Bekki2010}
{Bekki}, K., {Owers}, M.~S., and {Couch}, W.~J. 2010, \apjl, 718, L27.

\bibitem[{Bekki}, {Shioya} \& {Couch}(2001)]{Bekki2001}
{Bekki}, K., {Shioya}, Y., and {Couch}, W.~J. 2001, \apjl, 547, L17.

\bibitem[{Biviano} {\it et al.}\ (1997)]{Biviano1997}
{Biviano}, A. {\it et al.}\  1997, \aap, 321, 84.

\bibitem[{Biviano} {\it et al.}\ (2002)]{Biviano2002}
{Biviano}, A. {\it et al.}\  2002, \aap, 387, 8.

\bibitem[{Blake} {\it et al.}\ (2004)]{Blake2004}
{Blake}, C. {\it et al.}\  2004, \mnras, 355, 713.

\bibitem[{Blanton} {\it et al.}\ (2003)]{Blanton2003}
{Blanton}, M.~R. {\it et al.}\  2003, \apj, 594, 186.

\bibitem[{Boselli} \& {Gavazzi}(2006)]{Boselli2006}
{Boselli}, A. and {Gavazzi}, G. 2006, \pasp, 118, 517.

\bibitem[{Brinchmann} {\it et al.}\ (2004)]{Brinchmann2004}
{Brinchmann}, J. {\it et al.}\  2004, \mnras, 351, 1151.

\bibitem[{Butcher} \& {Oemler}(1984)]{Butcher1984}
{Butcher}, H.~R. and {Oemler}, Jr., A. 1984, \nat, 310, 31.

\bibitem[{Caldwell} \& {Rose}(1997)]{Caldwell1997}
{Caldwell}, N. and {Rose}, J.~A. 1997, \aj, 113, 492.

\bibitem[{Cava} {\it et al.}\ (2009)]{Cava2009}
{Cava}, A. {\it et al.}\  2009, \aap, 495, 707.

\bibitem[{Cohen}, {Hickox} \& {Wegner}(2015)]{Cohen2015}
{Cohen}, S.~A., {Hickox}, R.~C., and {Wegner}, G.~A. 2015, \apj, 806, 85.

\bibitem[{Cohen} {\it et al.}\ (2014)]{Cohen2014}
{Cohen}, S.~A. {\it et al.}\  2014, \apj, 783, 136.

\bibitem[{Cooper} {\it et al.}\ (2008)]{Cooper2008}
{Cooper}, M.~C. {\it et al.}\  2008, \mnras, 383, 1058.

\bibitem[{Couch} \& {Sharples}(1987)]{Couch1987}
{Couch}, W.~J. and {Sharples}, R.~M. 1987, \mnras, 229, 423.

\bibitem[{Cowie} {\it et al.}\ (1996)]{Cowie1996}
{Cowie}, L.~L. {\it et al.}\  1996, \aj, 112, 839.

\bibitem[{D'Onofrio} {\it et al.}\ (2014)]{DOnofrio2014}
{D'Onofrio}, M. {\it et al.}\  2014, \aap, 572, A87.

\bibitem[{Dressler}(1980)]{Dressler1980}
{Dressler}, A. 1980, \apj, 236, 351.

\bibitem[{Dressler} \& {Gunn}(1982)]{Dressler1982}
{Dressler}, A. and {Gunn}, J.~E. 1982, \apj, 263, 533.

\bibitem[{Dressler} \& {Gunn}(1992)]{Dressler1992}
{Dressler}, A. and {Gunn}, J.~E. 1992, \apjs, 78, 1.

\bibitem[{Dressler} {\it et al.}\ (1997)]{Dressler1997}
{Dressler}, A. {\it et al.}\  1997, \apj, 490, 577.

\bibitem[{Dressler} {\it et al.}\ (2013)]{Dressler2013}
{Dressler}, A. {\it et al.}\  2013, \apj, 770, 62.

\bibitem[{Dressler} {\it et al.}\ (1999)]{Dressler1999}
{Dressler}, A. {\it et al.}\  1999, \apjs, 122, 51.

\bibitem[{Ebeling} {\it et al.}\ (2000)]{Ebeling2000}
{Ebeling}, H. {\it et al.}\  2000, \mnras, 318, 333.

\bibitem[{Ebeling} {\it et al.}\ (1998)]{Ebeling1998}
{Ebeling}, H. {\it et al.}\  1998, \mnras, 301, 881.

\bibitem[{Ebeling} {\it et al.}\ (1996)]{Ebeling1996}
{Ebeling}, H. {\it et al.}\  1996, \mnras, 281, 799.

\bibitem[{Efron}(1982)]{Efron1982}
{Efron}, B. 1982, { {The Jackknife, the Bootstrap and other resampling plans}},
  ).

\bibitem[{Fasano} {\it et al.}\ (2010)]{Fasano2010}
{Fasano}, G. {\it et al.}\  2010, \mnras, 404, 1490.

\bibitem[{Fasano} {\it et al.}\ (2006)]{Fasano2006}
{Fasano}, G. {\it et al.}\  2006, \aap, 445, 805.

\bibitem[{Fasano} {\it et al.}\ (2015)]{Fasano2015}
{Fasano}, G. {\it et al.}\  2015, \mnras, 449, 3927.

\bibitem[{Fasano} {\it et al.}\ (2012)]{Fasano2012}
{Fasano}, G. {\it et al.}\  2012, \mnras, 420, 926.

\bibitem[{Finn} {\it et al.}\ (2005)]{Finn2005}
{Finn}, R.~A. {\it et al.}\  2005, \apj, 630, 206.

\bibitem[{Fritz} {\it et al.}\ (2007)]{Fritz2007}
{Fritz}, J. {\it et al.}\  2007, \aap, 470, 137.

\bibitem[{Fritz} {\it et al.}\ (2014)]{Fritz2014}
{Fritz}, J. {\it et al.}\  2014, \aap, 566, A32.

\bibitem[{Fritz} {\it et al.}\ (2011)]{Fritz2011}
{Fritz}, J. {\it et al.}\  2011, \aap, 526, A45.

\bibitem[{Gehrels}(1986)]{Gehrels1986}
{Gehrels}, N. 1986, \apj, 303, 336.

\bibitem[{Gill}, {Knebe} \& {Gibson}(2005)]{Gill2005}
{Gill}, S.~P.~D., {Knebe}, A., and {Gibson}, B.~K. 2005, \mnras, 356, 1327.

\bibitem[{Goto}(2004)]{Goto2004}
{Goto}, T. 2004, \aap, 427, 125.

\bibitem[{Goto}(2005)]{Goto2005}
{Goto}, T. 2005, \mnras, 357, 937.

\bibitem[{Goto} {\it et al.}\ (2003)]{Goto2003}
{Goto}, T. {\it et al.}\  2003, \pasj, 55, 771.

\bibitem[{Guglielmo} {\it et al.}\ (2015)]{Guglielmo2015}
{Guglielmo}, V. {\it et al.}\  2015, \mnras, 450, 2749.

\bibitem[{Gullieuszik} {\it et al.}\ (2015)]{Gullieusik2015}
{Gullieuszik}, M. {\it et al.}\  2015, \aap, 581, A41.

\bibitem[{Gunn} \& {Gott}(1972)]{Gunn1972}
{Gunn}, J.~E. and {Gott}, III, J.~R. 1972, \apj, 176, 1.

\bibitem[{Haines} {\it et al.}\ (2015)]{Haines2015}
{Haines}, C.~P. {\it et al.}\  2015, \apj, 806, 101.

\bibitem[{Haines} {\it et al.}\ (2013)]{Haines2013}
{Haines}, C.~P. {\it et al.}\  2013, \apj, 775, 126.

\bibitem[{Hern{\'a}ndez-Fern{\'a}ndez} {\it et al.}\
  (2014)]{Hernandez-Fernandez2014}
{Hern{\'a}ndez-Fern{\'a}ndez}, J.~D. {\it et al.}\  2014, \mnras, 438, 2186.

\bibitem[{Hogg} {\it et al.}\ (2006)]{Hogg2006}
{Hogg}, D.~W. {\it et al.}\  2006, \apj, 650, 763.

\bibitem[{Jaff{\'e}} {\it et al.}\ (2015)]{Jaffe2015}
{Jaff{\'e}}, Y.~L. {\it et al.}\  2015, \mnras, 448, 1715.

\bibitem[{Jaff{\'e}} {\it et al.}\ (2016)]{Jaffe2016}
{Jaff{\'e}}, Y.~L. {\it et al.}\  2016, \mnras, 461, 1202.

\bibitem[{Kauffmann} {\it et al.}\ (2003)]{Kauffmann2003}
{Kauffmann}, G. {\it et al.}\  2003, \mnras, 341, 54.

\bibitem[{Kauffmann} {\it et al.}\ (2004)]{Kauffmann2004}
{Kauffmann}, G. {\it et al.}\  2004, \mnras, 353, 713.

\bibitem[{Larson}, {Tinsley} \& {Caldwell}(1980)]{Larson1980}
{Larson}, R.~B., {Tinsley}, B.~M., and {Caldwell}, C.~N. 1980, \apj, 237, 692.

\bibitem[{Lewis} {\it et al.}\ (2002)]{Lewis2002}
{Lewis}, I. {\it et al.}\  2002, \mnras, 334, 673.

\bibitem[{Mahajan}(2013)]{Mahajan2013}
{Mahajan}, S. 2013, \mnras, 431, L117.

\bibitem[{Mahajan}, {Mamon} \& {Raychaudhury}(2011)]{Mahajan2011}
{Mahajan}, S., {Mamon}, G.~A., and {Raychaudhury}, S. 2011, \mnras, 416, 2882.

\bibitem[{Marziani} {\it et al.}\ (2016)]{Marziani2016}
{Marziani}, P. {\it et al.}\  2016, ArXiv e-prints.

\bibitem[{Mihos} \& {Hernquist}(1994)]{Mihos1994}
{Mihos}, J.~C. and {Hernquist}, L. 1994, \apjl, 425, L13.

\bibitem[{Mok} {\it et al.}\ (2013)]{Mok2013}
{Mok}, A. {\it et al.}\  2013, \mnras, 431, 1090.

\bibitem[{Moore} {\it et al.}\ (1996)]{Moore1996}
{Moore}, B. {\it et al.}\  1996, \nat, 379, 613.

\bibitem[{Moran} {\it et al.}\ (2007)]{Moran2007}
{Moran}, S.~M. {\it et al.}\  2007, \apj, 671, 1503.

\bibitem[{Moretti} {\it et al.}\ (2017)]{Moretti2017}
{Moretti}, A. {\it et al.}\  2017, ArXiv e-prints.

\bibitem[{Moretti} {\it et al.}\ (2014)]{Moretti2014}
{Moretti}, A. {\it et al.}\  2014, \aap, 564, A138.

\bibitem[{Muzzin} {\it et al.}\ (2014)]{Muzzin2014}
{Muzzin}, A. {\it et al.}\  2014, \apj, 796, 65.

\bibitem[{Newberry}, {Boroson} \& {Kirshner}(1990)]{Newberry1990}
{Newberry}, M.~V., {Boroson}, T.~A., and {Kirshner}, R.~P. 1990, \apj, 350,
  585.

\bibitem[{Oman}, {Hudson} \& {Behroozi}(2013)]{Oman2013}
{Oman}, K.~A., {Hudson}, M.~J., and {Behroozi}, P.~S. 2013, \mnras, 431, 2307.

\bibitem[{Omizzolo} {\it et al.}\ (2014)]{Omizzolo2014}
{Omizzolo}, A. {\it et al.}\  2014, \aap, 561, A111.

\bibitem[{Paccagnella} {\it et al.}\ (2016)]{Paccagnella2016}
{Paccagnella}, A. {\it et al.}\  2016, \apjl, 816, L25.

\bibitem[{Patel} {\it et al.}\ (2009)]{Patel2009}
{Patel}, S.~G. {\it et al.}\  2009, \apjl, 705, L67.

\bibitem[{Pimbblet} {\it et al.}\ (2002)]{Pimbblet2002}
{Pimbblet}, K.~A. {\it et al.}\  2002, \mnras, 331, 333.

\bibitem[{Poggianti}(2004)]{Poggianti2004}
{Poggianti}, B.~M. 2004, Clusters of Galaxies: Probes of Cosmological Structure
  and Galaxy Evolution, 245.

\bibitem[{Poggianti} {\it et al.}\ (2009)]{Poggianti2009}
{Poggianti}, B.~M. {\it et al.}\  2009, \apj, 693, 112.

\bibitem[{Poggianti} \& {Barbaro}(1996)]{Poggianti1996}
{Poggianti}, B.~M. and {Barbaro}, G. 1996, \aap, 314, 379.

\bibitem[{Poggianti} \& {Barbaro}(1997)]{Poggianti1997}
{Poggianti}, B.~M. and {Barbaro}, G. 1997, \aap, 325, 1025.

\bibitem[{Poggianti} {\it et al.}\ (1999)]{Poggianti1999}
{Poggianti}, B.~M. {\it et al.}\  1999, \apj, 518, 576.

\bibitem[{Poggianti} {\it et al.}\ (2006)]{Poggianti2006}
{Poggianti}, B.~M. {\it et al.}\  2006, \apj, 642, 188.

\bibitem[{Popesso} {\it et al.}\ (2007)]{Popesso2007}
{Popesso}, P. {\it et al.}\  2007, \aap, 461, 411.

\bibitem[{Quintero} {\it et al.}\ (2004)]{Quintero2004}
{Quintero}, A.~D. {\it et al.}\  2004, \apj, 602, 190.

\bibitem[{Ramella} {\it et al.}\ (2007)]{Ramella2007}
{Ramella}, M. {\it et al.}\  2007, \aap, 470, 39.

\bibitem[{Salpeter}(1955)]{Salpeter1955}
{Salpeter}, E.~E. 1955, \apj, 121, 161.

\bibitem[{Schawinski} {\it et al.}\ (2014)]{Schawinski2014}
{Schawinski}, K. {\it et al.}\  2014, \mnras, 440, 889.

\bibitem[{Smith} {\it et al.}\ (2004)]{Smith2004}
{Smith}, G.~A. {\it et al.}\  2004, in { Ground-based Instrumentation for
  Astronomy}, ed.\ A.~F.~M. {Moorwood} and M. {Iye}, volume 5492 of {
  \procspie}, 410.

\bibitem[{Taranu}, {Dubinski} \& {Yee}(2013)]{Taranu2013}
{Taranu}, D.~S., {Dubinski}, J., and {Yee}, H.~K.~C. 2013, \apj, 778, 61.

\bibitem[{Tran} {\it et al.}\ (2003)]{Tran2003}
{Tran}, K.-V.~H. {\it et al.}\  2003, \apj, 599, 865.

\bibitem[{Tran} {\it et al.}\ (2004)]{Tran2004}
{Tran}, K.-V.~H. {\it et al.}\  2004, in { IAU Colloq. 195: Outskirts of Galaxy
  Clusters: Intense Life in the Suburbs}, ed.\ A. {Diaferio}, 483.

\bibitem[{Tran} {\it et al.}\ (2007)]{Tran2007}
{Tran}, K.-V.~H. {\it et al.}\  2007, \apj, 661, 750.

\bibitem[{Treu} {\it et al.}\ (2003)]{Treu2003}
{Treu}, T. {\it et al.}\  2003, \apj, 591, 53.

\bibitem[{Valentinuzzi} {\it et al.}\ (2011)]{Valentinuzzi2011}
{Valentinuzzi}, T. {\it et al.}\  2011, \aap, 536, A34.

\bibitem[{Valentinuzzi} {\it et al.}\ (2009)]{Valentinuzzi2009}
{Valentinuzzi}, T. {\it et al.}\  2009, \aap, 501, 851.

\bibitem[{Varela} {\it et al.}\ (2009)]{Varela2009}
{Varela}, J. {\it et al.}\  2009, \aap, 497, 667.

\bibitem[{von der Linden} {\it et al.}\ (2007)]{vonderLinden2007}
{von der Linden}, A. {\it et al.}\  2007, \mnras, 379, 867.

\bibitem[{von der Linden} {\it et al.}\ (2010)]{vonderLinden2010}
{von der Linden}, A. {\it et al.}\  2010, \mnras, 404, 1231.

\bibitem[{Vulcani} {\it et al.}\ (2010)]{Vulcani2010}
{Vulcani}, B. {\it et al.}\  2010, \apjl, 710, L1.

\bibitem[{Vulcani} {\it et al.}\ (2015)]{Vulcani2015}
{Vulcani}, B. {\it et al.}\  2015, \apj, 798, 52.

\bibitem[{Weinmann} {\it et al.}\ (2006)]{Weinmann2006}
{Weinmann}, S.~M. {\it et al.}\  2006, \mnras, 366, 2.

\bibitem[{Wetzel} {\it et al.}\ (2013)]{Wetzel2013}
{Wetzel}, A.~R. {\it et al.}\  2013, \mnras, 432, 336.

\bibitem[{Wheeler} {\it et al.}\ (2014)]{Wheeler2014}
{Wheeler}, C. {\it et al.}\  2014, \mnras, 442, 1396.

\end{thebibliography}
\end{document}